\documentclass{article}
\usepackage{latexsym,natbib,amsmath,amssymb}
\usepackage{graphicx} %[dvips]
\usepackage{epstopdf,color}% To incorporate .eps illustrations using PDFLaTeX, etc.
\renewcommand{\vec}[1]{\mbox{\boldmath $#1$}} %vector
\def\dd{{\, \rm{d}}}
\def\dr{{\rm{d}}}

\def\bra{\langle}
\def\ket{\rangle}

\def\beq{\begin{equation}}
\def\eeq{\end{equation}}
\def\la{\label}
\def\ii{{\rm i}}

\def\r#1{(\ref{#1})}
\newcommand{\mylab}[3]{\raisebox{#2}[0mm][0mm]{%
\makebox[0mm][l]{\hspace*{#1}{#3}}}}%

\def\figpath{./}
\def\spacce#1{\hskip #1pt}
\def\drawline#1#2{\raise 2.5pt\vbox{\hrule width #1pt height #2pt}}
\def\solid{\drawline{24}{.5}\nobreak}

\def\bdash{\hbox{\drawline{5.8}{.5}\spacce{2}}}

\def\dashed{\bdash\bdash\bdash\nobreak}

\def\bdot{\hbox{\drawline{1}{.5}\spacce{2}}}

\def\dotted{\hbox{\leaders\bdot\hskip 24pt}\nobreak}

\def\chndot{\hbox%
{\drawline{4.6}{.5}\spacce{2}\drawline{1}{.5}\spacce{2}\drawline{4.6}{.5}\spacce{2}\drawline{1}{.5}\spacce{2}\drawline{4.6}{.5}}\nobreak }
\def\circle{$\circ$\nobreak }

\def\trian{\raise 1.25pt\hbox{$\scriptstyle\triangle$}\nobreak}

\def\dtrian{\raise 1.25pt\hbox%
{$\scriptscriptstyle\bigtriangledown$}\nobreak}

\def\squar{\raise 1.25pt\hbox{$\scriptstyle\Box$}\nobreak}

\def\diamon{\raise 1.25pt\hbox{$\scriptstyle\diamond$}\nobreak}

\newcommand{\soliddtrian}{$\blacktriangledown$\nobreak}

\def\linedtri1{\hbox{\bdash\hspace{-1.6mm}$\bigtriangleup$\hspace{-0.8mm}\bdash}\nobreak}
\def\soliddtrian1{$\blacktriangledown$\nobreak}
\def\solidrtrian2{$\blacktriangleright$\nobreak}
\def\solidltrian3{$\blacktriangleleft$\nobreak}

\title{Dipoles and streams in two-dimensional turbulence}

\author{Javier Jim\'enez\\
School of Aeronautics, U. Polit\'ecnica Madrid, 28040 Madrid Spain}
\date{\today}

\begin{document}
\maketitle
\begin{abstract}
Following the suggestion from the Monte--Carlo experiments in Jim\'enez, J. of Turbul.
(2020), that dipoles are as important to the dynamics of decaying two-dimensional turbulence
as individual vortex cores, it is found that the kinetic energy of this flow is carried by
elongated streams formed by the concatenation of dipoles. Vortices separate into a family of
small fast-moving cores, and another family of larger slowly moving ones, which can be
described as `frozen' into a slowly evolving `crystal'. The kinematics of both families are
very different, and only the former is self-similar. The latter is responsible for most of
the kinetic energy of the flow, and its vortices form the dipoles and the streams. 
Mechanisms are discussed for the growth of this slow component.
\end{abstract}

% ----------------------------------------
%\keywords{turbulence, segmentation, machine learning.}

%\linenumbers
% --------------------------------------------------------------------------
\section{Introduction}\la{sec:intro}

The subject of this paper originates from the Monte-Carlo simulations in
\cite{jimploff18,jimploff20,jotploff}, whose purpose was to identify causally significant
structures in two-dimensional decaying turbulence by inspecting the effect, after some
predetermined time, of randomly perturbed initial conditions. These experiments will not be
repeated here, and the reader is directed to the original publications for details. One of the most
interesting results was that, besides the expected identification of individual vortices as
significant \citep{mcwilliams84,mcwilliams90}, the experiments found that tight dipoles of
counterrotating vortices are as causally important as isolated vortices, or even more
so. Modifying a strong vortex in the initial conditions leads to a large perturbation of the
flow after five to ten eddy turnovers, but modifying a dipole leads to an even stronger
perturbation. Corrotating vortex pairs were not found to be significant in the same way.

The experiments mentioned above were intended to validate the Monte Carlo procedure, as well
as to answer the fundamental question of whether some localised flow regions are more
important than others for the evolution of the flow. They did not pay too much attention to
the properties of the flow itself, being restricted, among other things, to a single
relatively low Reynolds number. The present paper deals with the fluid mechanics. In
particular, it examines whether our understanding of the evolution of two-dimensional
turbulence can be improved by the consideration of collective structures, such as the
dipoles mentioned above.

Several questions need to be addressed. The first one has to do with the Reynolds number,
because one of the results in \cite{jotploff} was that the preferred scale for significant vortices
and dipoles is about the same, even if the vortices are associated with the enstrophy and
the dipoles are structures of the kinetic energy. Spectral analysis shows that the typical
scales associated with these two variables are different, but the Reynolds number in
\cite{jotploff} was not high enough to separate them clearly. To clarify this
question, we analyse flow simulations at several Reynolds numbers, allowing,  at
least, for some range of scales.

The second question has to do with the role played in the flow by the two `templates'
(vortices and dipoles), and perhaps by other structures, because the original analysis was
not concerned with the best representation of the flow, nor with flow mechanisms. Its only
purpose was to identify which structures are most important from the point of view of
dynamics, but not to clarify the dynamics itself. This problem is also connected with the
Reynolds number, because flows at low Reynolds numbers essentially contain a single scale,
which represents everything. More general flows are multiscale, and it is usually true that
structures that represent well some aspect the flow are not the ones that control the
dynamics of others. For example, although vortices and vortex stretching \citep{vinc91} are
considered good models for the turbulence energy cascade \citep{rich20,betc56}, at least
from the point of view of enhancing dissipation, it was shown by \cite{jwsr} that removing
them from the flow had very little lasting effect, and there is clear evidence of
intermediate scales of the kinetic energy that are involved in the cascade process without
being directly related to vorticity \citep{cardesa17}.
  
Much of the interest in two-dimensional turbulence originates from the remark by
\cite{onsag} that the inviscid evolution of a high-energy system of point vortices results
in negative temperature states, and that this would naturally lead to the formation of
organised coherent structures, rather than to a disordered flow.

There are at least two ways of approximating high-Reynolds-number two-dimensional turbulence
by a conservative Hamiltonian system. The first one is the aforementioned system of point
vortices \citep{bat67}, and the second is the approximation of the inviscid Euler equations
in terms of Fourier components of the velocity, truncated to a finite range of 
wavenumber magnitude \citep{Basd:Sad:75,les97}.

\cite{kraichnan67} followed the suggestion of \cite{onsag} to propose that forced two-dimensional turbulence
should include a reverse energy cascade towards larger scales, as well as a direct enstrophy
cascade towards smaller ones. There is a fair agreement on the mechanism of the enstrophy
cascade by means of vortex amalgamation and filamentation
\citep{mcwilliams90,carnevale91,Benzi92,dritsch08}. The inverse cascade is less well understood,
although it is generally believed that its mechanics is different from that of the enstrophy
cascade and, in particular, that it is not predominantly mediated by vortex merging
\citep{Par:Tab:98,Boff:Cel:Ver:00,eyink06,xiao:09}. 

\cite{kraichnan67} derived the form of the spectrum of the truncated equilibrium Euler
system, and observed that, in the absence of a low-wavenumber dissipation mechanism, energy
would tend to accumulate at the largest system scale, in a process similar to the
Bose--Einstein condensation of quantum systems. Both the reverse cascade and the condensate
\citep{smithyak93,smithyak94} have been numerically and experimentally observed. A fair
amount of work has gone into finding equilibrium solutions of the Euler equations that could
account for this long-term flow behaviour, from vortex crystals \citep{aref02} to maximum
entropy statistics \citep{Joy:Mont:73,Mont:Joy:74}, and there is numerical evidence that
forced viscous two-dimensional flow locally relax to these inviscid equilibrium solutions
\citep{Mont:Etal:92,Mont:Shan:Matth:93}. For example, forced numerical turbulence in a
square box evolves to a large-scale dipole filling the box diagonally
\citep{smithyak93}. Recent reviews of the existing work on the reverse energy cascade and on
condensate states can be found in \cite{tabeling02} and \cite{Boff:Eck:12}.

Most of the work on the two-dimensional energy cascade has used forced experiments in which
the system eventually settles to a statistically steady state. This has the advantage of
allowing the use of ergodicity to compile statistics, but complicates the interpretation of
the results because of the constant interference from the forcing. Decaying turbulence also
has an inverse energy flux to large scales (although not necessarily an inverse cascade,
as we will see later in the paper). The total energy remains approximately constant
while its length scale grows until it collides with the domain size. From the point of view
of causality characterisation, decaying turbulence has obvious advantages, because things
happen only once, and the arrow of time is well defined. On the negative side, compiling
reliable statistics requires ensembles of simulations, and the analysis of the resulting large
data sets. The final stage of Bose condensation is also of little interest in this
case, because the flow decays before it can be completed, but we will see in
\S\ref{sec:collective} that a related process is important in the creation of large-scale
structures.

In this paper, we study the dynamics of two-dimensional decaying
turbulence, with emphasis on the mechanics of the inverse energy flux, using ensembles of
simulations at low to moderate Reynolds numbers, guided by the results of the causality
analysis mentioned at the beginning of this introduction.
The simulations are described in \S \ref{sec:simul} , followed in \S \ref{sec:templates} by
the structural analysis of the flow in terms of the vortices, dipoles and streams suggested
by the causal analysis. Section \ref{sec:collective} describes the collective organisation
of the vortices, including their classification into types and how each type is related to
the large scales of the kinetic energy. Section \ref{sec:tur2d} discusses mechanisms for this
organisation, and \S\ref{sec:conc} concludes.

%\newpage 

% --------------------------------------------------------------------------------------
\section{Simulations and basic flow properties}\la{sec:simul}

% =======================================================
\begin{table}
  \begin{center}
    \def~{\hphantom{0}}
    \begin{tabular}{lcccccccccccc}
       Case  & $N$ & $L_{init}/L$ & $q'_0L/\nu$  & $\lambda_{\tau 0}/L$ & 
              $\lambda_{\omega 0}/L$ & $\lambda_{5 0}/L$ &   
              $Re_{\lambda 0}$  &$\omega'_0t_F$ & $\omega'_F/\omega'_0$ & $q'_F/q'_0$ & Symbol \\[3pt]
      T256 & 256 & 0.1 & $2300$ & 0.044  & 0.184 & 0.058 &100 & 3.3  & 0.73 & 0.95 & \circle \\
      T512 & 512 & 0.05 & $4400$ & 0.030  & 0.145 & 0.036 & 132 & 18.4  & 0.69 & 0.95 & \trian \\
      T768 & 768 & 0.025 & $7800$ & 0.025  & 0.116 & 0.024 &  197 & 33.7 & 0.69 & 0.96 & \dtrian \\
      T1024 & 1024 & 0.033 & $11000$ & 0.022 & 0.105 & 0.019 & 250& 44.9 & 0.69 & 0.96 & \squar \\
    \end{tabular}
\caption{Parameters of the simulations. The size of the doubly periodic computational box is $L\times
L$. The r.m.s. vorticity $\omega_0'$, and the velocity magnitude $q_0'$, are measured after the
initial discarded transient, decaying from an initial enstrophy spectrum whose peak is at
wavelength $L_{init}$. The Taylor microscale, $\lambda_{\tau 0}= q_0'/\omega'_0$ is used to
compute the Reynolds number $Re_\lambda=q'_0\lambda_{\tau 0}/\nu$, and $\lambda_{\omega 0}$
and $\lambda_{50}$ are the enstrophy and palinstrophy scales, respectively, defined in
figure \ref{fig:specs}. The number of collocation points used before dealiasing is $N\times
N$. The subscript `F' refers to the end of each simulation, so that the decay time after the
initial transient is $t_F$. Case T256 was used in \cite{jotploff} to identify the causally
significant structures used as starting points for the discussion in the text. Each case is
an ensemble of 768 independent experiments.
 }
    \label{tab:cases}
  \end{center}
\end{table}
% =======================================================

% ===========================================================
\begin{figure}
%\vspace*{8mm}%
\centerline{%
\raisebox{0mm}{\includegraphics[height=0.315\textwidth,clip]{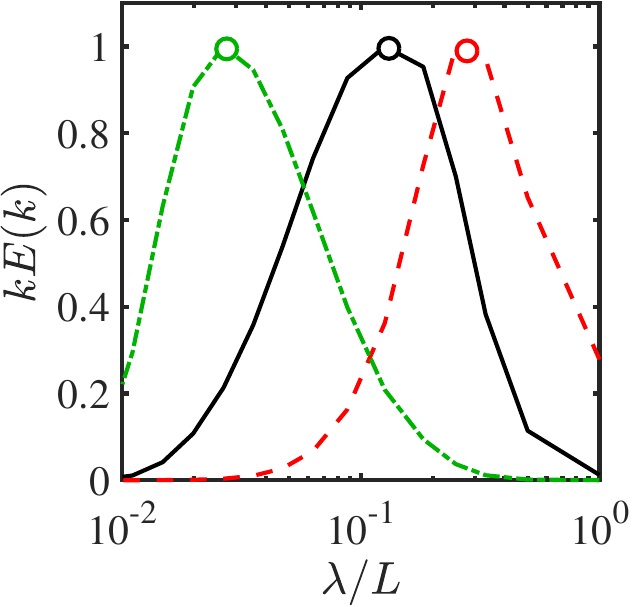}}%spwithpeaks}}%
\mylab{-.26\textwidth}{.27\textwidth}{(a)}%
\hspace*{2mm}%
\raisebox{1mm}{\includegraphics[height=0.305\textwidth,clip]{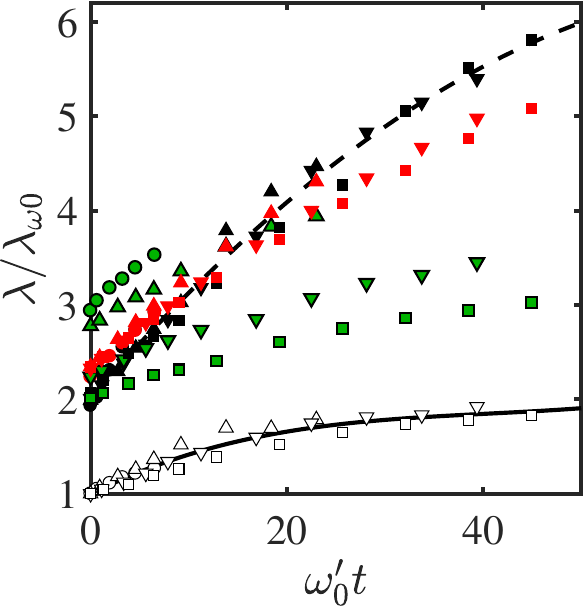}}%sppeaks}}%
\mylab{-.24\textwidth}{.27\textwidth}{(b)}%
\hspace*{2mm}%
\raisebox{0.5mm}{\includegraphics[height=0.31\textwidth,clip]{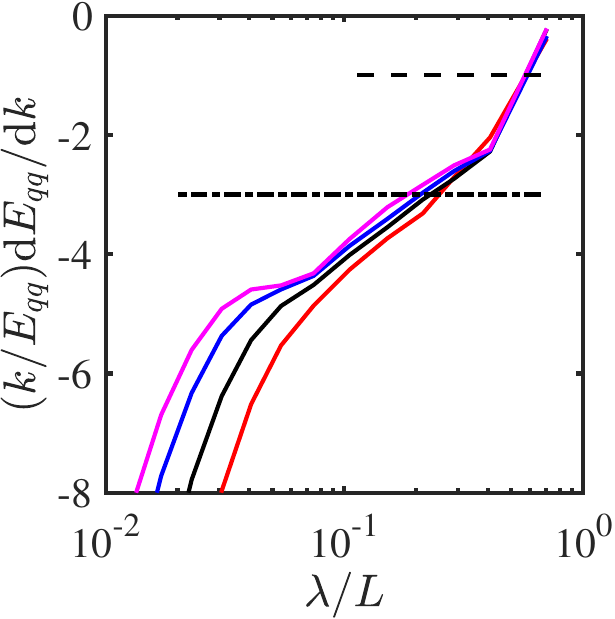}}%spslope}}%
\mylab{-.25\textwidth}{.27\textwidth}{(c)}%
}
\vspace{4mm}
\centerline{%
\raisebox{0mm}{\includegraphics[height=0.31\textwidth,clip]{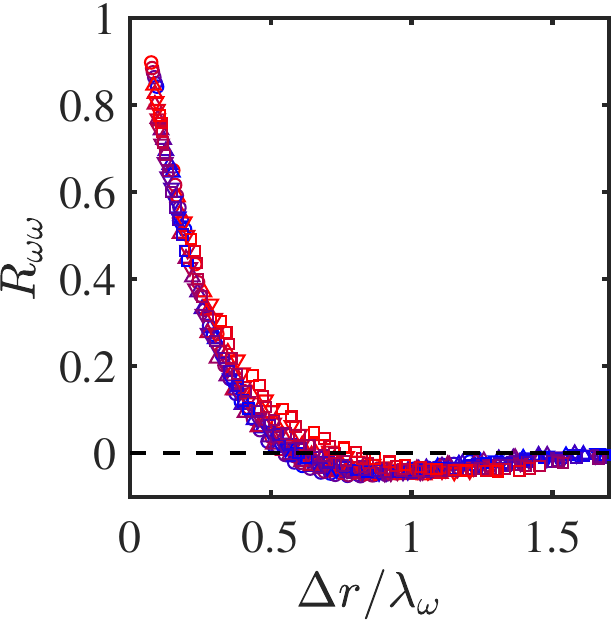}}%DeltaomIN}}%
\mylab{-.05\textwidth}{.27\textwidth}{(d)}%
\hspace*{2mm}%
\raisebox{0mm}{\includegraphics[height=0.31\textwidth,clip]{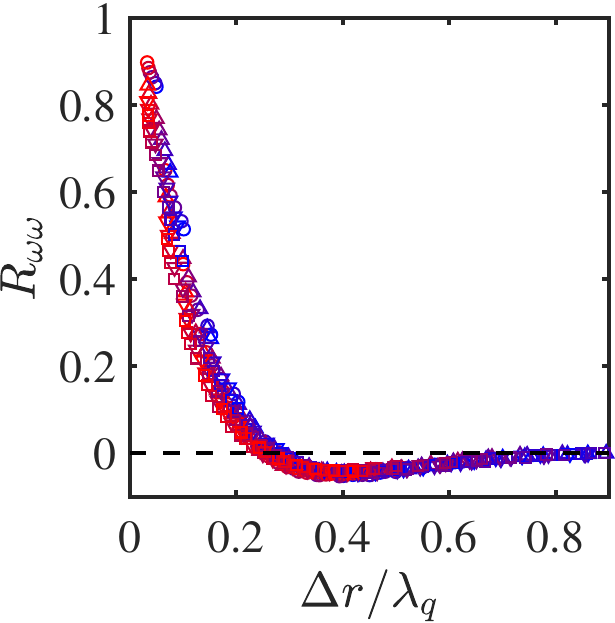}}%DeltaomOUT}}%
\mylab{-.05\textwidth}{.27\textwidth}{(e)}%
\hspace*{2mm}%
\raisebox{0mm}{\includegraphics[height=0.31\textwidth,clip]{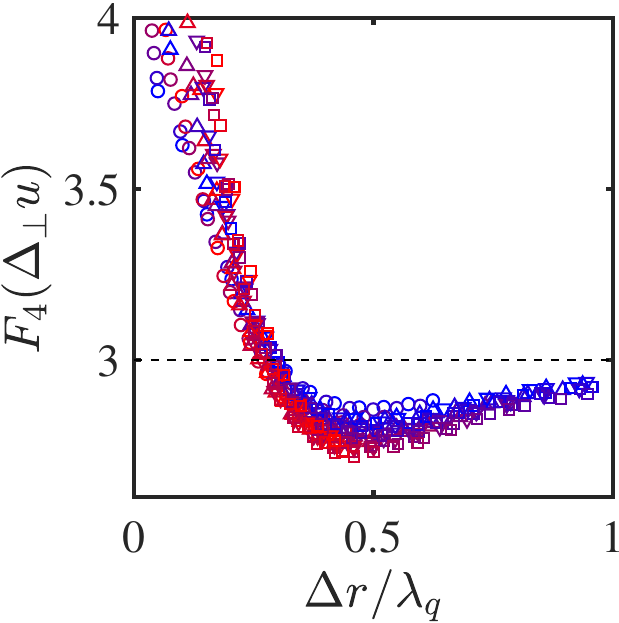}}%Fu4}}%
\mylab{-.05\textwidth}{.27\textwidth}{(f)}%
}
\caption{%
(a) Definition of the enstrophy, palinstrophy, and energy peak wavelengths. \solid, Premultiplied enstrophy
spectrum; \dashed, energy spectrum; \chndot, palinstrophy.  Case T768 at $\omega_0' t =6.3$.
(b) Evolution of the enstrophy and energy peak wavelengths; normalized with the enstrophy
wavelength at $t=0$. Symbols as in table \ref{tab:cases}. Open symbols are enstrophy, and
closed black ones are energy. The two polynomial fits are used as reference in later figures.
The red closed symbols are the Taylor scale, stretched for clarity to $10\lambda_\tau/\lambda_{\omega 0}$,
and the green ones are the stretched palinstropy wavelength,  $10\lambda_5/\lambda_{\omega 0}$; 
(c) Logarithmic slope of the energy spectrum at the end of each simulation. From left to
right, T1024 to T256. The two horizontal lines mark slopes $-1$ and $-3$, which respectively
correspond to the energy and enstrophy peak wavelengths in (a).
(d) Correlation function of the vorticity, as a function of the separation, $\Delta r$,
scaled with $\lambda_\omega$. Symbols as in table \ref{tab:cases}. Colours label time along
each simulation for a fixed $\Delta r$, from $t=0$ in blue to $t_F$ in red.
(e) As in (d), scaled with $\lambda_q$. 
(f) Fourth-order flatness factor for the transverse velocity increments, $\Delta_\perp u= u(x,y+\Delta r)-u(x,y)$. Symbols
and colours as in (d). 
}
\label{fig:specs}
\end{figure}
% ===========================================================

Simulations of decaying nominally isotropic two-dimensional turbulence are performed at
various scale disparities in a doubly periodic square box of side $L$, using a standard
spectral Fourier code dealiased by the 2/3 rule. Time advance is third-order Runge-Kutta.
The flow field is defined by its velocity $\vec{u}=(u,v)$ in the plane $\vec{x}=(x,y)$, and
by the one-component vorticity $\omega=\nabla\times\vec{u}$. It is initialised with random
Fourier phases and a fixed isotropic enstrophy spectrum, which is relatively flat for small
wavenumber magnitude $k$, $E_{\omega\omega}\sim k^{3/2}$, and much steeper,
$E_{\omega\omega}\sim k^{-25/2}$, for large ones. The peak of this initial spectrum, located
at $k_{init} \approx 2\pi/L_{init}$, controls the initial energy-containing spectral range.
The simulations solve the Navier--Stokes equations in vorticity--stream-function
formulation, using regular second-order viscosity, $\nu\nabla^2\omega$.

Natural time and velocity scales can be defined from the root-mean-square (r.m.s.) vorticity
magnitude $\omega' =\bra \omega^2\ket ^{1/2}$, where $\bra\cdot\ket$ is the time-dependent
ensemble average taken over the full computational box, and from $q' =(u'^2 + v'^2)^{1/2}$.
The flow is allowed to evolve for $q' t_{init}/L=0.32$ $(\omega't_{init}\approx
6\mbox{--}12)$, after which the structures have established themselves. This moment is
defined as the start of the simulations, $t=0$, for the rest of the paper, and is denoted by
a `0' subindex in the corresponding quantities. After some experimentation, the evolution of
most quantities of interest for our argument was found to collapse best when plotted against
$\omega'_0 t$, which will be used in the following. However, this collapse is not universal,
and we could not find any normalisation for the time that would collapse well the evolution
of all the flow properties, including some basic ones like the decay of the kinetic energy
and of the enstrophy. As the simulation proceeds, the enstrophy decays by approximately
50\%, while the kinetic energy decreases at most by 5--10\%. A Taylor length scale can be
defined as $\lambda_\tau=q'/\omega'$, and used to define a microscale Reynolds number,
$Re_\lambda=q'\lambda_\tau/\nu$, where $\nu$ is the kinematic viscosity. Both grow by
factors of 1.5--2.5 during each simulation, depending on the simulation time. Finally, each
experiment is repeated at least 768 times to compile statistics, and a few cases were
repeated twice as many times, to test convergence. These parameters are summarised in table
\ref{tab:cases}.

\begin{figure}
\vspace*{4mm}%
\centering
\raisebox{0mm}{\includegraphics[height=0.31\textwidth,clip]{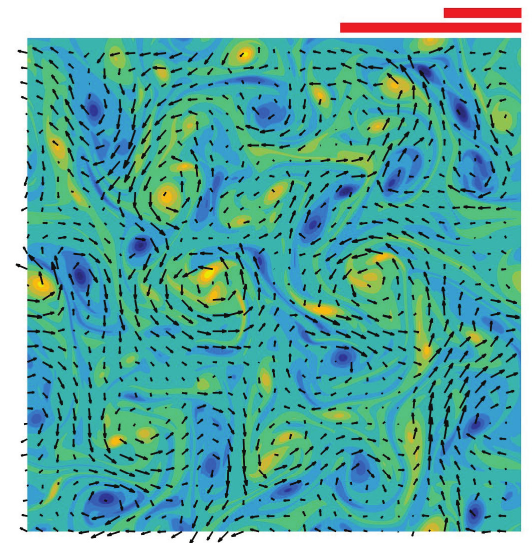}}%noploff1024_t50_0_phys_it10.png}}%
\mylab{-.17\textwidth}{.32\textwidth}{(a)}%
\hspace*{2mm}%
\raisebox{0mm}{\includegraphics[height=0.31\textwidth,clip]{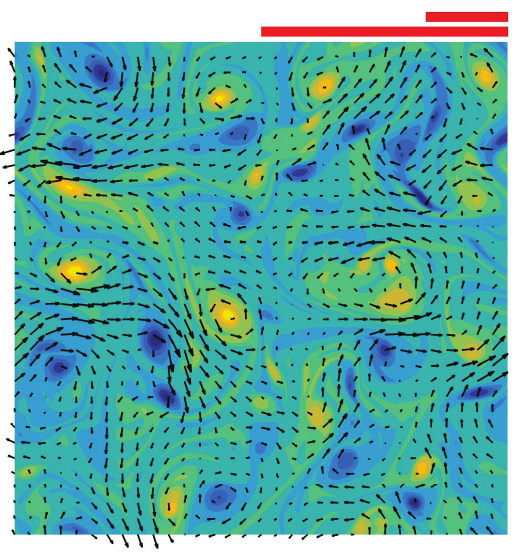}}%noploff1024_t50_0_phys_it20.png}}%
\mylab{-.17\textwidth}{.32\textwidth}{(b)}%
\hspace*{2mm}%
\raisebox{0mm}{\includegraphics[height=0.31\textwidth,clip]{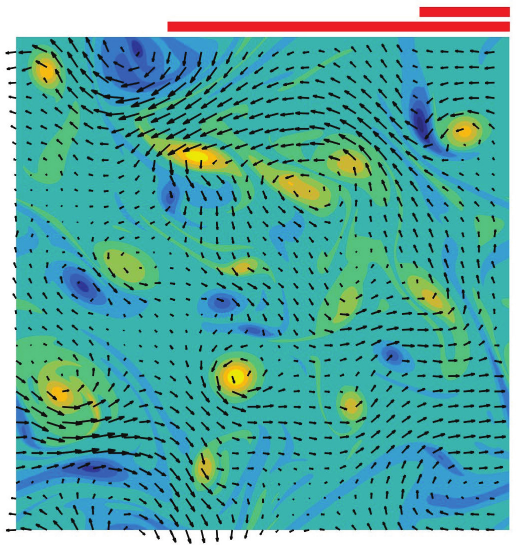}}%noploff1024_t50_0_phys_it50.png}}%
\mylab{-.17\textwidth}{.32\textwidth}{(c)}%
\caption{%
(a) Vorticity and velocity field of a typical flow from T1024. From left to right:
$\omega'_0 t=12.8$, 25.6, 64.2. The shorter bar on top of each figure is $\lambda_\omega$.
The longer one is $\lambda_q$.
}
\label{fig:omfields}
\end{figure}

The evolution of the energy and enstrophy spectra is displayed in figure \ref{fig:specs}.
For each simulation, time-dependent length scales for the vorticity and for the velocity can
respectively be defined by the location of the maximum of the premultiplied enstrophy and
energy spectra, as illustrated in figure \ref{fig:specs}(a). The enstrophy wavelength,
$\lambda_\omega=2\pi/k_\omega$, increases only slowly with time, as shown in figure
\ref{fig:specs}(b), but the energy scale, $\lambda_q$, increases faster as the energy flows
towards larger sizes. This is also clear from the snapshots in figure \ref{fig:omfields},
which include scale bars for both wavelengths. Eventually, $\lambda_q\approx L$, at which
moment the reverse energy flux saturates, and $\lambda_q$ stops growing \citep{smithyak93}.
Although all our simulations were originally run for $\omega_0' t\approx 60$, only times for
which $\lambda_q/L<0.6$ are included in figure \ref{fig:specs}(b) and in table
\ref{tab:cases}. The flow is considered to enter afterwards into a different
energy-condensation phase of its evolution, which is not discussed in this paper. Figure
\ref{fig:specs}(c) displays the logarithmic slope of the energy spectrum. A plateau in this
figure represents a power-law range, and the figure shows that no such range develops in our
simulations at the level of $k^{-3}$, corresponding to the classical enstrophy cascade
\citep{kraichnan67,batchelor69}. Inspection of the temporal evolution of the spectrum (not
shown) suggests that a weak inflection point may be developing at this level, but, since the
spectra in figure \ref{fig:specs}(c) are plotted towards the end of each simulation, it is
unlikely that a $k^{-3}$ range would ever develop in them. A slope $(k/E_{qq})\dr E_{qq}/\dr
k=-3$ corresponds (algebraically) to the maximum of $k E_{\omega\omega}=k^3 E_{qq}$, which
was used above to define $\lambda_\omega$. An extended $k^{-3}$ range would correspond to a
flat top of the premultiplied enstrophy spectrum in figure \ref{fig:specs}(a). This does not
happen, and the premultiplied enstrophy spectrum stays relatively sharp, as in figure
\ref{fig:specs}(a). In the same way, a wavelength $\lambda_5$ can be defined by the maximum
of the premultiplied spectrum of the vorticity gradient, $kE_{\nabla\omega\nabla\omega}= k^5
E_{qq}$ (figure \ref{fig:specs}a). The magnitude of this gradient is sometimes called the
`palinstrophy', and controls the viscous dissipation of the enstrophy \citep{les97}. We will
therefore call $\lambda_5$ the palinstrophy wavelength. It is 3--5 times smaller than
$\lambda_\omega$, and its initial value is given in table \ref{tab:cases}. It is the true
viscous dissipative length for the flow, and the shortest of all the length scales defined
here.

If we take $\lambda_5$ to be a measure of the smallest vorticity structures, the
numerical resolution of the simulations in table \ref{tab:cases} is $\Delta x
/\lambda_5=0.1-0.2$, in terms of complex Fourier modes, and improves as the simulations
proceed.

A logarithmic slope of $-1$ coincides with the energy wavelength, and, because all the spectra in
figure \ref{fig:specs}(c) are drawn just before the growth of $\lambda_q$  saturates in their
respective simulations, $\lambda_q/L\approx 0.5$ in all cases.

% ===========================================================
% ===========================================================

It is interesting that the spectra in figure \ref{fig:specs}(c) develop a short but clear
$k^{-5}$ power law around the dissipative wavelength $\lambda_5$, especially for the higher
Reynolds numbers. Spectra steeper than $k^{-3}$ are well-known in two-dimensional
turbulence, and are believed to originate from a variety of reasons that depend on the
initial conditions. \cite{saff71} argued that, since vorticity is conserved in the inviscid
two-dimensional limit, the mutual distortion of the vortices eventually leads to the
formation of vorticity discontinuities whose spectrum is $E_{\omega\omega}\sim k^{-2}$, or
$E_{qq}\sim k^{-4}$. \cite{brach88}, starting from a relatively smooth vorticity field,
report that a $k^{-4}$ energy spectrum appears initially, but evolves into $k^{-3}$ after
individual vortices appear. This initial time ($\omega'_0 t\approx 10$) is of the same order
as the transient period discarded in our simulations.
   
On the other hand, \cite{mcwilliams84} and \cite{Benzi87}, whose initial conditions already
include a $k^{-3}$ spectral range, develop a steeper slope in the later part of the decay
($\omega'_0 t\gtrsim 100$ in our notation), when most of the vorticity is organised into
individual cores that approximately behave as a conservative Hamiltonian system
\citep{bat67}. They relate this steeper spectral slope to the vorticity distribution in
the cores. This seems to be the case in our simulations, where the $k^{-5}$ plateau only
develops towards the energy-condensed end of each simulation. We will mostly be interested
in the earlier part of the evolution, where both the vortex cores and the `incoherent'
background vorticity are relevant. It should be noted that most of the simulations by the
authors mentioned above use high-order hyperviscosity, which favours the formation of
isolated vortices, instead of the regular viscosity in this paper. 

Figure \ref{fig:specs}(d,e) display the autocorrelation function for the vorticity,
\beq
R_{\omega\omega}=\bra \omega(x,y)\omega(x+\Delta r,y)\ket/\omega'^2.
\la{eq:Rww}
\eeq
It has an inner core that scales with $\lambda_\omega$ in figure \ref{fig:specs}(d),
and an outer region that scales with $\lambda_q$ in figure \ref{fig:specs}(e).
The scaling in figure \ref{fig:specs}(d) suggest that the diameter of the vortices
that contribute to the correlation is $O(0.3\lambda_\omega)$. This will be confirmed when
we study individual cores in \S\ref{sec:vortices}. At the moderate Reynolds numbers and
regular viscosity of our simulations, the flow velocity is only slightly intermittent. The
fourth-order flatness of the transversal velocity increments across a distance $\Delta r$ is
shown in figure \ref{fig:specs}(f). It also has two distance ranges. In an intermittent inner core
of the order of the vortex diameter, $\Delta r/\lambda_\omega \approx 0.3$,
the flatness reaches $F_4\approx 4$. This part of the distribution does not collapse well in
the figure, which is drawn  to emphasise the return to
gaussianity at larger distances. The outer part of the distribution scales with $\lambda_q$.
The flatness decays to the slightly sub-gaussian value $F_4\approx 2.8$ at $\Delta r/
\lambda_q\approx 0.4$, and relaxes to the gaussian $F_4\approx 3$ beyond $\Delta r\approx
\lambda_q$. The velocity itself is always close to gaussian. Since the velocity gradient is
known to be very intermittent near individual vortex cores, where it is dominated by the
$1/r$ behaviour of the velocity \citep{jimenez96}, the return to gaussianity marks the distance
at which the flow is dominated by the interaction among several cores, instead of
by individual ones. Figure \ref{fig:specs}(f) thus suggests that this typical distance among
strong cores is $O(\lambda_q)$. This agrees with the outer scaling of figure
\ref{fig:specs}(e), and will also be confirmed in \S\ref{sec:vortices}.

The Taylor microscale and the viscous length $\lambda_5$ are  included in
figure \ref{fig:specs}(b), on a stretched vertical scale for clarity. Note that,
although $\lambda_q$ and $\lambda_\tau$ collapse among the different simulations when
normalised with $\lambda_{\omega 0}$, the viscous length $\lambda_5$ does not. The reason is
uncertain, but relatively unimportant for the present paper, which is  concerned with the
large scales of the flow rather than with  small-scale features. Viscous enstrophy
dissipation is often described as being preceded by filamentation of the vortex cores
under mutual stretching \citep{dritsch08}, and $\lambda_5$ probably measures the width of
these filaments and of the vortex fragments into which they decay.

 It should also be noted that, although the Taylor scale is probably little
more than an arbitrary length scale in three-dimensional turbulence, it has a deeper
significance in two dimensions. We mentioned in the introduction the coexistence of two
cascades in two-dimensional turbulence, and that they are connected in a general way with
the form of the equilibrium spectrum of the truncated Fourier representation of the Euler
equations. The only parameter in this spectrum is the ratio between the total kinetic energy
and the enstrophy, which is the squared Taylor scale, and it can be shown that the limiting
wavelength between the two cascades is proportional to $\lambda_\tau$
\citep{Basd:Sad:75,les97}. Whether an inverse cascade exists depends on whether this
limit falls within the truncated set of Fourier wavenumbers, or, equivalently, on the ratio
$\lambda_\tau/L$. It is difficult to make this criterion quantitative in viscous flows,
whose spectrum is very far from equilibrium, but the growth of this ratio with time in
figure \ref{fig:specs}(b) signals a shift of the kinetic energy towards larger scales.

% ---------------------------------------------------------------------------------------------------
\section{Structural models}\la{sec:templates}

% ===========================================================
\begin{figure}
%\vspace*{4mm}%
\centering
\raisebox{0mm}{\includegraphics[height=0.35\textwidth,clip]{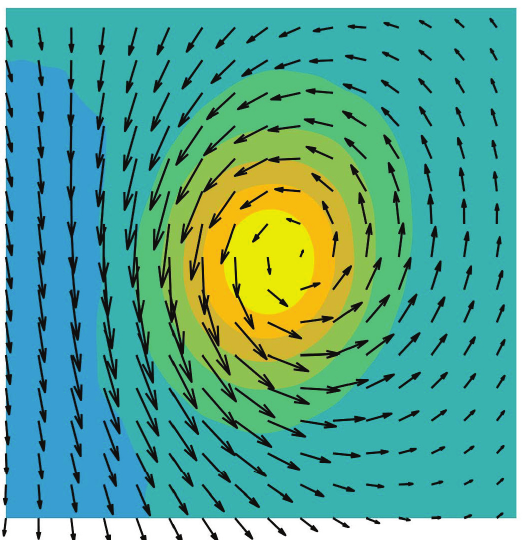}}%maxtemp1.png}}%
\mylab{-.38\textwidth}{.30\textwidth}{(a)}%
\hspace*{10mm}%
\raisebox{1mm}{\includegraphics[height=0.35\textwidth,clip]{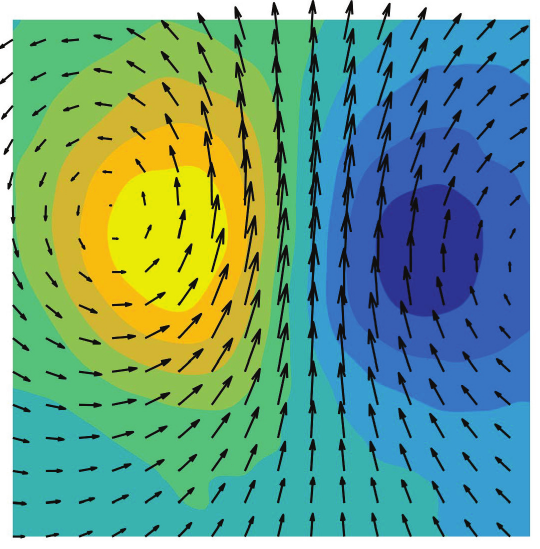}}%maxtemp6.png}}%
\mylab{-.39\textwidth}{.30\textwidth}{(b)}%
\caption{%
Vorticity and velocity field for the templates identified in \cite{jotploff} as the most causally
relevant flow features. (a) Vortex template, mostly relevant for experiments manipulating
vorticity. (b) Dipole template, relevant for velocity manipulations. Because the position,
scale and intensity of the templates are adjusted when matching the flow, their orientation,
size and intensity are arbitrary.
}
\label{fig:temp}
\end{figure}
% ===========================================================

We saw in the introduction that the most interesting structural result in \cite{jotploff} is
that the causally significant flow neighbourhoods in two-dimensional turbulence look either
like isolated vortices, or like counterrotating vortex dipoles. The former was expected,
because the classical model for this flow is a vortex `gas'
\citep{mcwilliams84,mcwilliams90,carnevale91,Benzi92}, but the latter was a mild surprise.
The two models were summarised in \cite{jotploff} by the archetypal `templates' shown in
figure \ref{fig:temp}. They were obtained by conditionally averaging the flow
patches found to be most causally significant in the experiments described in the
introduction. The patch size was found to be important, and, after some experimentation, it was
adjusted to maximise the difference between significant and non-significant neighbourhoods.
The optimum choice at the Reynolds number of \cite{jotploff} are square patches of side
$L_p/L=0.1$, and the templates are constructed of side $L_T=3L_p$ to include some surrounding
flow. But there is no reason to assume that this `most causal' scale is also the one at
which templates optimally represent the flow, or that the same dimensions work at different
Reynolds numbers. \cite{jotploff} showed that the first question could be answered by a
posteriori optimisation of the approximation error between scaled templates and test flow
fields. The application of this procedure to the present flows, and the answer to the second
question, are tested below. Note that, when templates are used as archetypal flow
structures, they lose their original connection with causal significance. They are treated
as flow features that have been found to be of interest `for some reason', and the issue
becomes to find what they are, and whether they can be used as indicators of some aspect of
the flow dynamics. Why they were causally significant in the first place can be addressed a
posteriori, if desired. For more details on the process of template extraction, verification
and validation, the reader is directed to \cite{jimploff18,jotploff}.

To test how well a template approximates a particular flow neighbourhood, it is first scaled
to size $L_T\times L_T$, optimally oriented (using four orthogonal rotations and one
reflection), and its intensity is adjusted to match the overall r.m.s. intensity of the flow
in question, so that $\bra \xi_T^2\ket_T =\xi'^2$. The `T' subindex here represents template
properties, as well as averaging over the template domain, and $\xi$ is either the vector
velocity or the scalar vorticity. Centring the rescaled template at some point, $\vec{x}$,
the representation error is measured as the relative $L_2$ norm of the difference between
the flow and the template,
\beq
\Phi_\xi (\vec{x},L_T) = \frac{\|\xi(\vec{x}+\vec{\tilde{x}})-\xi_T(\vec{\tilde{x}})\|_{\vec{\tilde{x}}\in T}}{\|\xi\|_T},
\la{eq:temp1}
\eeq
which is a function of $L_T$ and of $\vec{x}$. Statistics
are compiled over all the positions in an $N_t\times N_t$ `test' grid, and over all the flow
realisations, and scanned over $L_T$.

% ===========================================================
\begin{figure}
%\vspace*{1mm}%
\centerline{%
\raisebox{0mm}{\includegraphics[height=0.35\textwidth,clip]{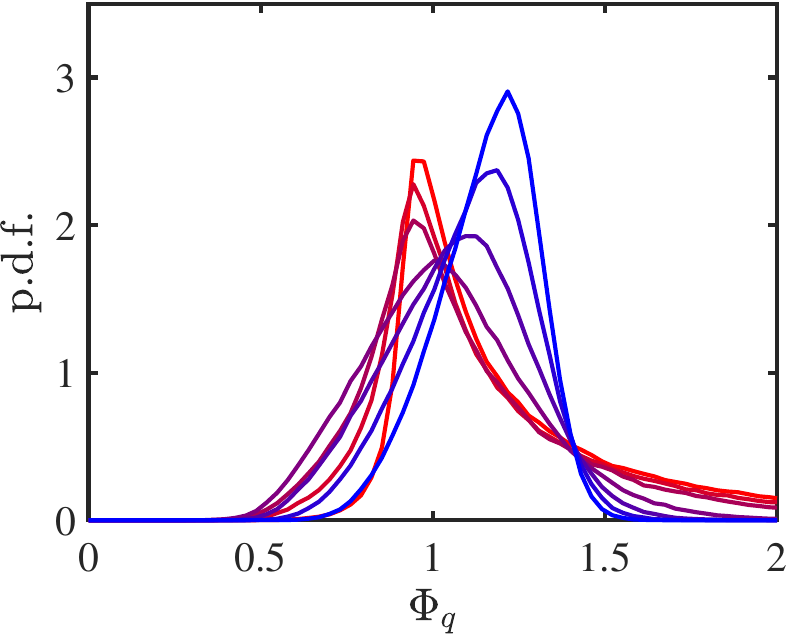}}%errboxv_sp2_tim6-q}}%
\mylab{-.35\textwidth}{.30\textwidth}{(a)}%
\hspace*{11mm}%
\raisebox{0mm}{\includegraphics[height=0.35\textwidth,clip]{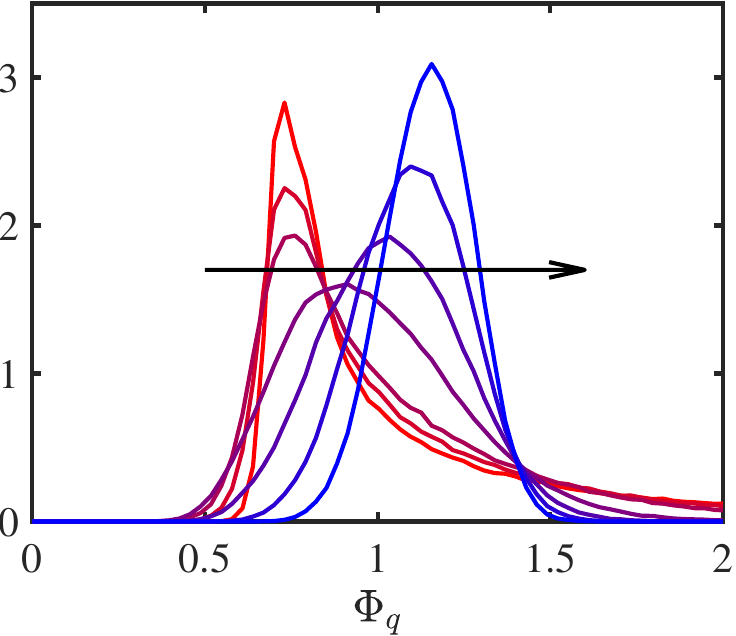}}%errboxd_sp2_tim6-q}}%
\mylab{-.35\textwidth}{.30\textwidth}{(b)}%
}%
\vspace{2mm}%
\centerline{%
\raisebox{0mm}{\includegraphics[height=0.35\textwidth,clip]{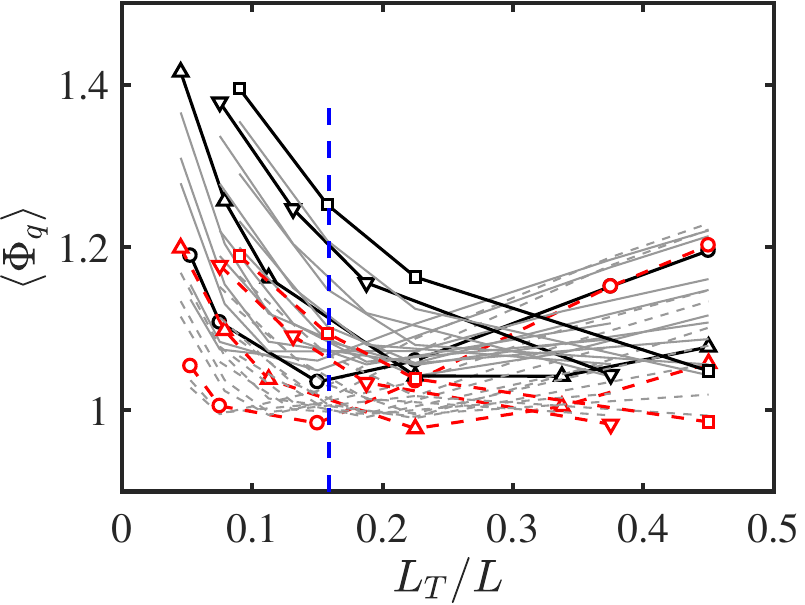}}%errvslt_av}}%
\mylab{-.34\textwidth}{.315\textwidth}{(c)}%
\hspace*{4mm}%
\raisebox{0mm}{\includegraphics[height=0.355\textwidth,clip]{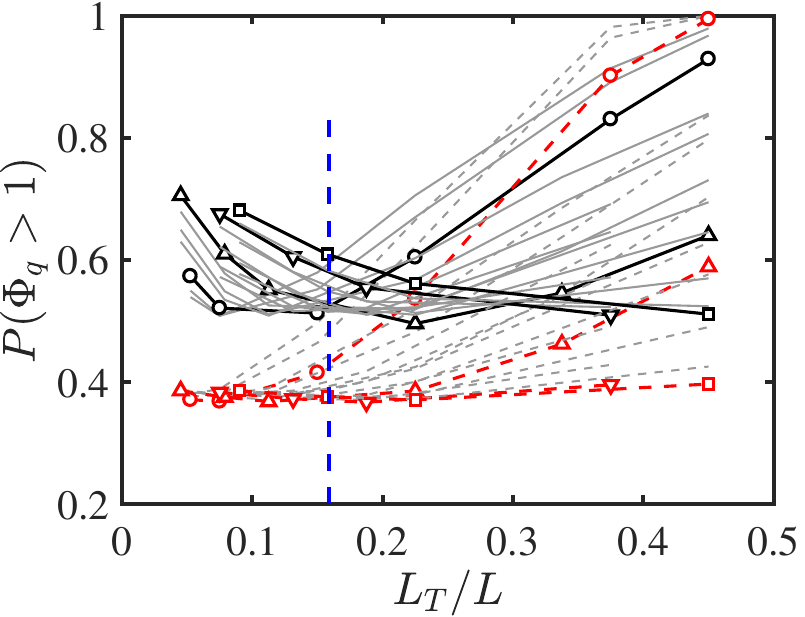}}%errvslt_f10}}%
\mylab{-.34\textwidth}{.312\textwidth}{(d)}%
}%
\vspace{2mm}%
\centerline{%
\raisebox{0mm}{\includegraphics[height=0.35\textwidth,clip]{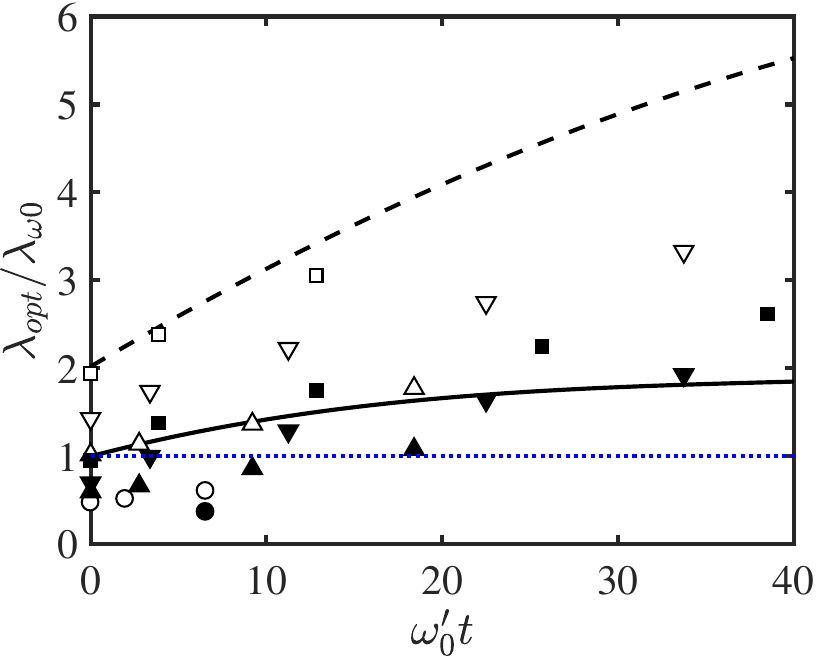}}%lamopt_f10_q}}%
\mylab{-.34\textwidth}{.315\textwidth}{(e)}%
\hspace*{5mm}%
\raisebox{0mm}{\includegraphics[height=0.35\textwidth,clip]{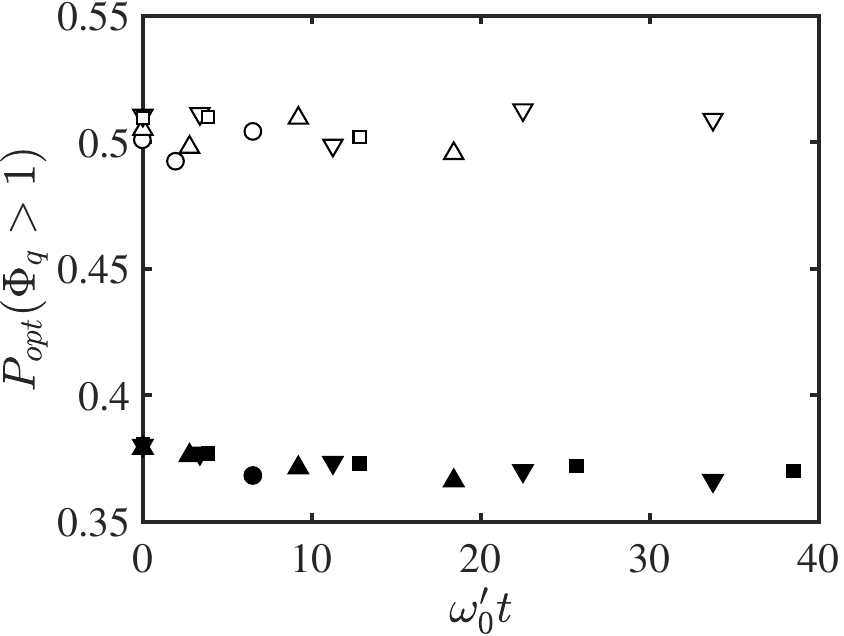}}%erropt_f10_q}}%
\mylab{-.34\textwidth}{.315\textwidth}{(f)}%
}
\caption{%
(a,b) P.d.f. of the template approximation error. Energy error norm. Case T512, $\omega'_0
t=9.2$. Template size, increasing from red to blue: $L_T/L=0.045$, 0.08, 0.11, 0.22, 0.34,
0.45, 0.56. (a) Template is a vortex. (b) Template is a dipole. The arrow is in the
direction of increasing  template size, $L_T$.
(c,d) Approximation error as a function of case and of $L_T$. Cases are plotted for
different times as grey lines without labels, except for the final time of each simulation,
which is highlighted and labelled as in table \ref{tab:cases}. \solid, Template is a
vortex; \dashed, template is a dipole. The dashed vertical line is a representative value of
$\lambda_\omega/L$, from table \ref{tab:cases}.
(c) Error is averaged over all template positions.
(d) Error measured as the fraction, $P_q$, of relative local errors larger than unity.
(e) Template size for optimum $P_q$. Lines are the polynomial fits to $\lambda_\omega$ and
$\lambda_q$ in figure \ref{fig:specs}(b). Closed symbols are dipole templates; open symbols
are vortex templates. (f) As in (e), showing the optimum $P_q$.
}
\label{fig:errhist}
\end{figure}
% ===========================================================

Typical probability density functions (p.d.f.s) of the approximation error are given in
figure \ref{fig:errhist}(a,b) as functions of the template size. The peak of the histogram
generally moves to smaller errors as $L_T$ decreases, and becomes very skewed, especially
for dipoles, suggesting the necessity of using several figures of merit to quantify the
overall performance of a template. An obvious choice is to minimise the mean error, $\bra \Phi_\xi
\ket(L_T)$, where the average is taken over all the template positions. It provides an
overall goodness of fit, but the shape of the histograms in figure \ref{fig:errhist}(a,b)
suggests that it may mix some very good local fits with some very bad ones. Coherent
structures can be relevant to the flow dynamics even when they fill a relatively small area
fraction \citep{jim18}, and a representation of the flow in terms of them should be able to
stress the good fits even at the expense of de-emphasizing some of the bad ones. A measure
with this property is the fraction of the p.d.f. above a given error threshold. We use
$P_{\xi}=\mbox{prob.}(\Phi_\xi>1)$. The behaviour of both measures with template size is
shown in figure \ref{fig:errhist}(c,d), and depends on the particular case and on the
simulation time considered. Most cases are displayed as light grey lines,
without identification, to show general trends, but the final time of each simulation,
defined by $\lambda_q/L\approx 0.6$, is highlighted and labelled with the symbols in table
\ref{tab:cases}.

At short simulation times (grey lines), figure \ref{fig:errhist}(c) shows that the average error, $\bra
\Phi_q \ket$, is minimum for template sizes of the order of $\lambda_\omega$, but that the
optimum size increases with time and with the Reynolds number. The optimum template size for some
of the longest simulations is the largest one allowed by the computational box,
$L_T/L\approx 0.45$. Larger templates are considered in this paper to be contaminated by the
box size, and are not included in the analysis. This behaviour holds for vortices (solid
lines), and for dipoles (dashed lines).

Figures \ref{fig:errhist}(d) and \ref{fig:errhist}(e) show that the fraction, $P_{q}$, of
large kinetic-energy error tends to be minimised by small templates of the order of the
diameter of the intense vortices seen in figure \ref{fig:omfields}. This fits the intent of
this measure, which is to identify local structures. It is interesting that this
preference for small features is clearest for the dipole template, while the optimum size
for the vortex template tends to be larger. Both optimum sizes increase with the evolution
time and with the Reynolds number. Figure \ref{fig:errhist} uses the kinetic-energy norm.
The results for the enstrophy norm are similar, with a preference for slightly larger
templates.

Figure \ref{fig:errhist}(f) shows that the approximation error attained at the respective
optimum size changes little among simulations and among simulation times, even if we have
seen that the optimum size varies widely. The optimum mean error is approximately $\bra
\Phi_q\ket =1.04,\, \bra \Phi_\omega\ket=1.22$, for vortices and $\bra \Phi_q\ket = 0.98,\,
\bra \Phi_\omega\ket=1.12$, for dipoles. The minimum error fraction is
$P_{q}=0.51,\,P_{\omega}=0.81$ for vortices and $P_{q}=0.37,\,P_{\omega}=0.62$, for dipoles.
Interestingly, dipoles are always more successful templates than isolated vortices, and the
kinetic-energy error is always lower than the enstrophy one. We will almost exclusively
discuss dipole templates and the kinetic-energy norm for the rest of this section.

% ---------------------------------------------------------------------------------------------------
\subsection{Large-scale streams.}\la{sec:large}

% ===========================================================
\begin{figure}
\vspace*{4mm}%
\centering
\raisebox{0mm}{\includegraphics[height=0.30\textwidth,clip]{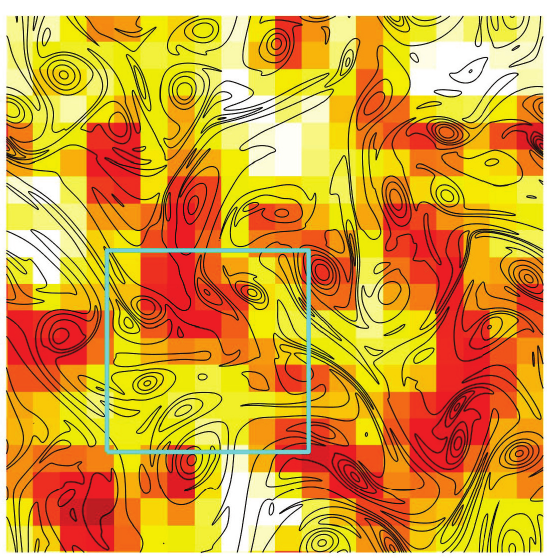}}%heat_v_f3t4f10.png}}%
\mylab{-.17\textwidth}{.32\textwidth}{(a)}%
\hspace*{5mm}%
\raisebox{-1mm}{\includegraphics[height=0.31\textwidth,clip]{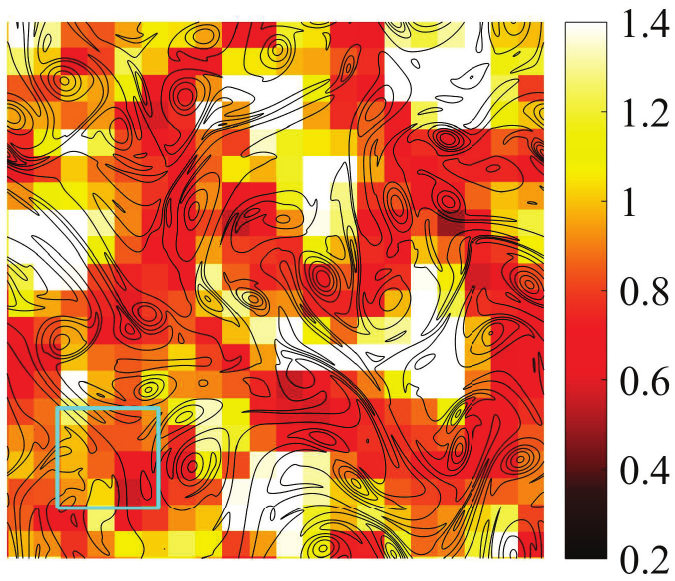}}%heat_d_f3t4f10.png}}%
\mylab{-.22\textwidth}{.32\textwidth}{(b)}%
\hspace*{1mm}%
\raisebox{0mm}{\includegraphics[height=0.315\textwidth,clip]{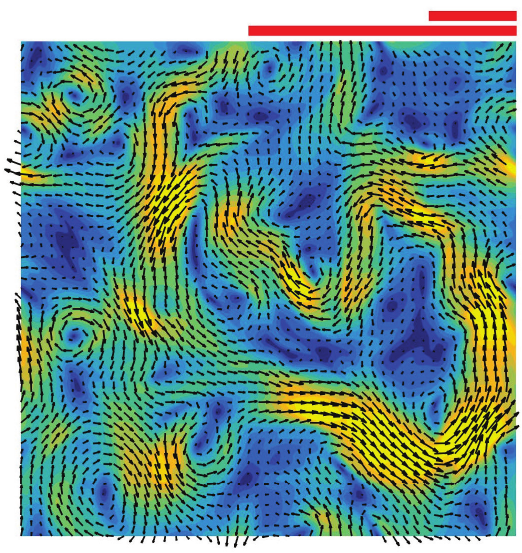}}%noploff768_t50_0_onlyu_it20.png}}%
\mylab{-.17\textwidth}{.32\textwidth}{(c)}%
\caption{%
(a,b) Heat maps for the fit error of a typical flow field. Case T768 at $\omega'_0 t=22$.
Colour represents the approximation error, $\Phi_q$, for a template centred at each point.
Line contours are the vorticity magnitude. The cyan box is the size of the template,
chosen in each case to optimise $P_q$. (a) Vortex template. (b) Dipole template.
(c) Velocity field for the flow in (a,b). The scale bars above the figure are as in figure
\ref{fig:omfields}. The background colour is the velocity magnitude, lighter for faster
velocities.
}
\label{fig:heat}
\end{figure}
% ===========================================================

Figures \ref{fig:heat}(a,b) are heat maps for the fit of the vortex and dipole templates to
a typical flow snapshot. The intensity in a heat map is proportional to the goodness of fit
of the template centred at that point. The darker regions in figure \ref{fig:heat} represent
better fits, measured by $\Phi_q$. Figure \ref{fig:heat}(a) is drawn for vortex templates,
and figure \ref{fig:heat}(b) is drawn for dipoles. In both cases, the size of the template
is chosen to be optimum, which in this particular case is different for the two templates
(see the cyan boxes in the figure). Inspection of the figures shows that the
features extracted by the two templates have much in common, although they differ in detail.
This is not surprising because a dipole is formed by two vortices, and it was to be expected
that at least some of the vortices identified by a vortex template are part of a dipole.

However, the main use of heat maps is not to identify individual features, but to highlight
the organisation of the features themselves. Thus, although we have mentioned that figures \ref{fig:heat}(a) and
\ref{fig:heat}(b) differ in detail, it is visually clear that they cluster around a common
large-scale structure, present in both figures. Its nature is clearer in figure
\ref{fig:heat}(c) which shows that it is a meandering stream that spans the full box. This
is interesting for two reasons. The first one is that it suggests that the feature 
detected by the dipole template is not the pair of vortices, but the jet between them.
The second one is that those jet segments are part of a larger stream, too large and too
irregular to be represented by any local template, but which can be recovered by the
concatenation of several of them.

% ===========================================================
\begin{figure}
\vspace*{5mm}%
\centerline{%
\raisebox{0mm}{\includegraphics[height=0.34\textwidth,clip]{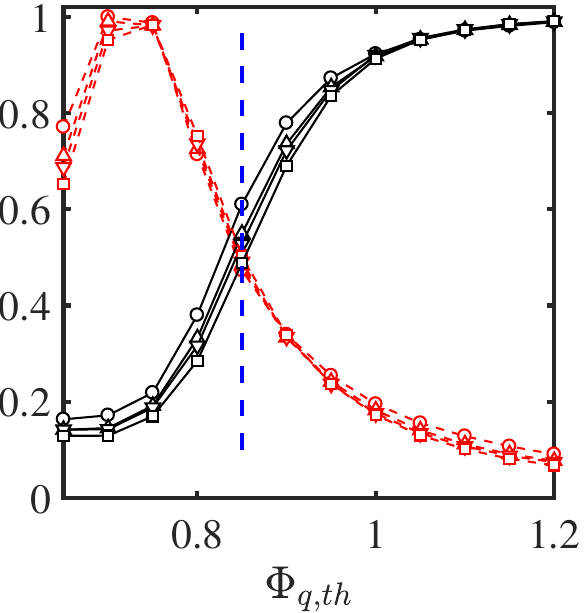}}%percol}}%
\mylab{-.17\textwidth}{.36\textwidth}{(a)}%
\hspace*{8mm}%
\raisebox{7.5mm}{\includegraphics[height=0.303\textwidth,clip]{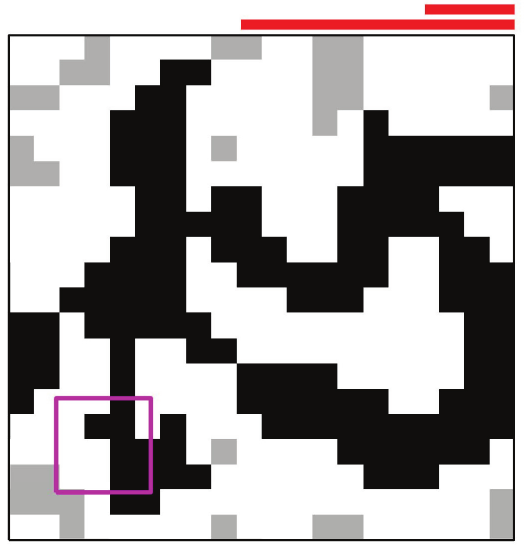}}%heath_d_f3t4_th85f10.png}}%
\mylab{-.17\textwidth}{.36\textwidth}{(b)}%
\hspace*{5mm}%
}%
\caption{%
Definition of the largest thresholded structure in each snapshot.
(a) Percolation diagram for $\Phi_q\le \Phi_{q,th}$, as a function of the threshold. Dipole
templates optimised for $P_{q}$. Solid lines are the area of the largest thresholded object
divided by the total thresholded area. Dashed lines are the number of objects, normalised to
unit maximum. Individual lines are for each simulation, with symbols as in table
\ref{tab:cases}. The vertical dashed line is the standard threshold used below.
(b) Thresholded heat map from figure \ref{fig:heat}(b). The darker object is the largest
connected structure, used in the following to represent the large-scale flow organisation,
and the purple box is the template size.
}
\label{fig:heathdef}
\end{figure}
% ===========================================================

Figure \ref{fig:heathdef} shows how this large-scale flow organisation can be defined by
thresholding the heat map below a given error level. The resulting points are collected into
individual objects, defined by contiguity along the four directions of the coordinate axes.
Figure \ref{fig:heathdef}(a) is the percolation diagram \citep{moisy04}. The solid lines
represent the fraction of the total thresholded area contained in the largest contiguous
thresholded object. It is unity for very high thresholds, where a single object fills the
whole field, and also for a lowest limit in which a single point represents both the whole
thresholded region and its largest object. Neither limit gives information about the structure
of the flow, and both are left outside figure \ref{fig:heathdef}(a), which centres on the
intermediate range in which several individual low-error objects first appear and then merge
into larger ones as the threshold is raised. The dashed lines are the number of individual
objects, normalised to unit maximum. The percolation diagram is averaged over all the times
of each simulation, and varies little among simulations. After some experimentation, the
reference threshold is chosen to be $\Phi_{q,th}=0.85$, which is used in figure
\ref{fig:heathdef}(b) to threshold the map in figure \ref{fig:heat}(b). There is a dark
largest connected object, and several smaller ones in a lighter colour, which are also below
the error threshold but which are not connected to the largest one. They will not be used
when compiling the statistics of the large flow scales. The percolation transition is
narrow, and changing the threshold by any large amount moves the result into either an empty
or a completely full map, but thresholds in the range 0.8--0.9 yield approximately the same
results as those presented below.

\begin{figure}
\vspace{5mm}
\centerline{%
\raisebox{0mm}{\includegraphics[height=0.32\textwidth,clip]{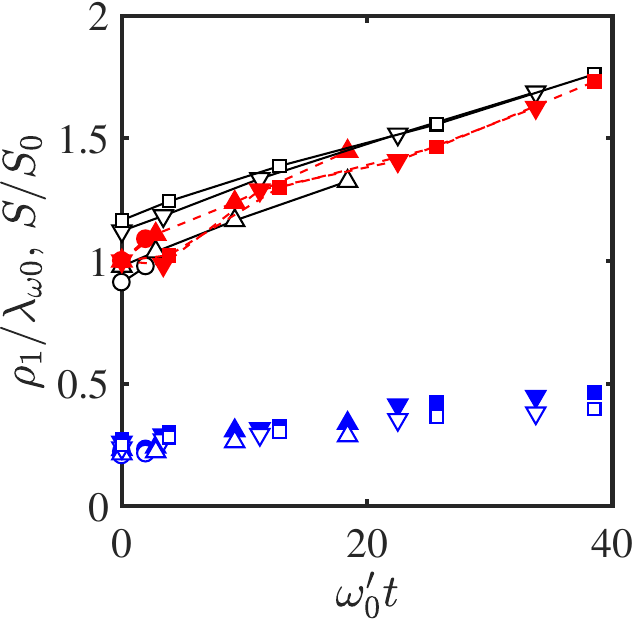}}%r1r2_th85_f10_r1S}}%
\mylab{-.16\textwidth}{.33\textwidth}{(a)}%
\hspace{1mm}%
\raisebox{0mm}{\includegraphics[height=0.32\textwidth,clip]{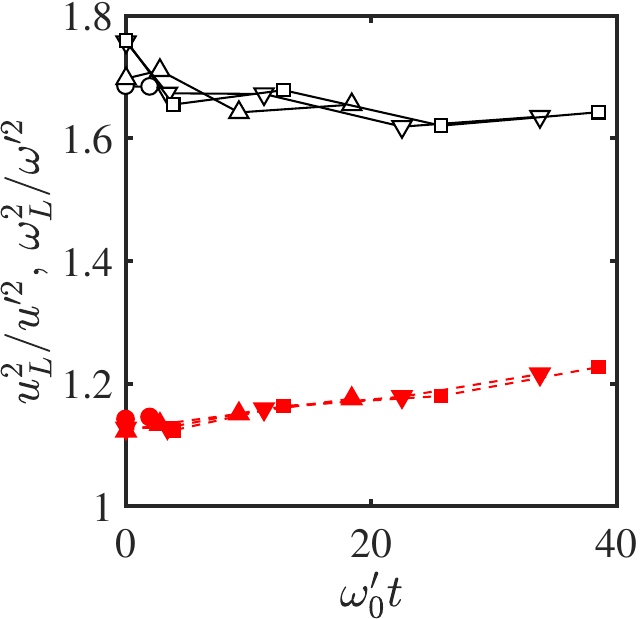}}%r1r2_th85_f10_uomL}}%
\mylab{-.16\textwidth}{.33\textwidth}{(b)}%
\hspace{1mm}%
\raisebox{0mm}{\includegraphics[height=0.32\textwidth,clip]{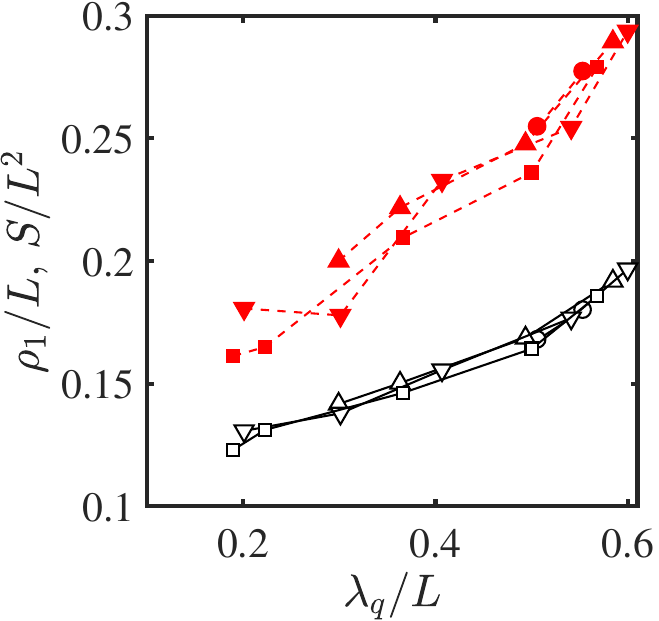}}%r1r2_th85_f10_rqvslq}}%
\mylab{-.16\textwidth}{.33\textwidth}{(c)}%
}%
\caption{%
Properties of the largest thresholded structure  of low $P_q$ in each snapshot.
(a) Temporal evolution of the geometry of the largest structure. \solid, Inner
scale normalised with the initial enstrophy scale, $\rho_1/\lambda_{\omega 0}$; \dashed,
area of the largest structure normalised with its initial value, $S/S_0$; symbols without
lines are the integral length $L_{int}/\lambda_{\omega 0}$ for the correlations of $\Phi_q$,
defined in \r{eq:Lint}. Symbols as in table \ref{tab:cases}. Open symbols are vortex
templates; closed symbols are dipoles.
(b) Flow properties within the largest structures. \solid, kinetic-energy density; \dashed, enstrophy density. 
(c) As in (a), but unnormalised,  versus the kinetic-energy wavelength.  
}
\label{fig:heath}
\end{figure}

It should be understood that heat maps and their thresholded versions are at most
`skeletons' of flow properties. Each point of the map is an element of the $N_t\times N_t$
`test' grid used to test the template approximation properties. It marks the centre of a
template box, but the optimum template size is generally wider than the spacing of the test
grid, as shown in figures \ref{fig:heat} and \ref{fig:heathdef}. Any geometric property of
the skeletons should be interpreted with this in mind. For example, the solid lines in
figure \ref{fig:heath}(a) show the inner `width', $\rho_1$, of the largest thresholded
object in each frame, defined as the side of the largest square that completely fits within
the object \citep{cat:dim:96,moisy04}. The figure is compiled over a test grid with
$N_t=20$, so that the minimum possible value is $\rho_1/L=0.05$. This is smaller than the
widths in figure \ref{fig:heath}(a), which are of the order of $\lambda_{\omega0}\approx
0.15$ (table \ref{tab:cases}), but close enough to it to recommend testing whether $\rho_1$
is influenced by the test grid. Limited testing with $N_t=30$ shows that the results in
figure \ref{fig:heath}(a,b) could change by approximately 15--20\% on a much finer test
grid.
  
% ===========================================================
% ===========================================================

The symbols without lines in figure \ref{fig:heath}(a) are the integral length,
\beq
L_{int}= \int_0^\infty C_{\Phi\Phi}(r) \dd r,
\la{eq:Lint}
\eeq
derived from the radial autocorrelation function of the approximation error, $\Phi_q$, which
typically measures the narrowest dimension of the structures of that variable. It is
interesting that, even if we have seen that the approximation error is a marker for the
largest, energy-containing scales in the flow, the integral length is narrower than the
enstrophy peak in figure \ref{fig:specs}. In fact, its typical value, $0.3\lambda_{\omega
0}\approx 0.05 L$, is of the order of the spacing, $L/N_t$ of the test grid over which it is
computed, and could actually be shorter in a finer grid. If we take $L_{int}$ to represent
the average thickness of the structures of $\Phi_q$, it would imply that the snapshot in figure
\ref{fig:heathdef}(b), where the thickness of the structures is of the order of
the test cell, is indeed typical.

Note that, since the length of the structures can be estimated by $S/\rho_1$, the fact that
$S$ and $\rho_1$ grow at a similar rate in figure \ref{fig:heath}(a) implies that the length
of the structures changes little as the flow evolves $(S/\rho_1L=1.3\to 1.5)$, but that their
aspect ratio grows `fatter' $(S/\rho_1^2=11\to 8)$. We will see below that the longitudinal
scale of the structures is typically of the order of the box size, and therefore
presumably limited by it.

Figure \ref{fig:heath}(b) shows kinetic energy and enstrophy averaged over dipole templates
centred on points within the largest thresholded object in each snapshot. As suggested by
figure \ref{fig:heat}(c), and by the definition of the dipole template in figure
\ref{fig:temp}(b), dipoles contain locally high kinetic energy, but also moderately high
enstrophy. Interestingly, enstrophy is less concentrated than the energy. It could be
expected that, since the edge of a jet is necessarily the seat of high vorticity, the
`fringe' of grid points surrounding the large-scale thresholded streams would be regions of especially
high enstrophy, but this is not true. When energy and enstrophy are conditioned to that
fringe, the kinetic energy density is lower than the average, and the enstrophy density is
indistinguishable from the mean (not shown). A similar result holds for other flow regions
outside the streams. The only high-energy regions are apparently those detected by the
dipole template, including the lighter-grey regions in figure \ref{fig:heathdef}(b), and the
only moderately-high enstrophy regions are also, on average, associated with the streams.

Perhaps the strongest indication of the connection of the streams with the structure of the
kinetic-energy is figure \ref{fig:heath}(c). The quantities in figure \ref{fig:heath}(a) are
normalised with their initial value because they do not otherwise collapse. The initial
conditions, which are chosen to provide a variety of scale combinations, are too different
to allow it. For example, the average value of the area of the largest stream at $t=0$
varies by 60\% among the different simulations. On the other hand, figure
\ref{fig:heath}(c) shows the unnormalised area and thickness of the large streams versus
the energy wavelength, $\lambda_q$. They collapse well, strongly suggesting that the streams
are the support of the kinetic energy of the flow, and that the process of stream formation is
tantamount to the flux of the energy to larger scales.

% ===========================================================
\begin{figure}
\vspace*{5mm}%
\centerline{%
\raisebox{4mm}{\includegraphics[height=0.27\textwidth,clip]{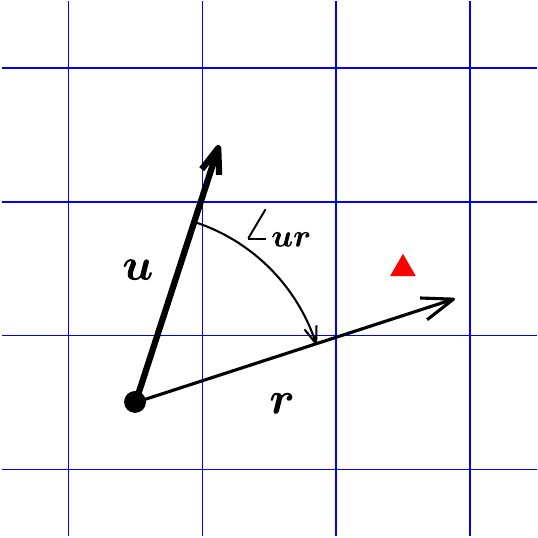}}%anglesketch}}%
\mylab{-.16\textwidth}{.32\textwidth}{(a)}%
\hspace*{5mm}%
\raisebox{0mm}{\includegraphics[height=0.30\textwidth,clip]{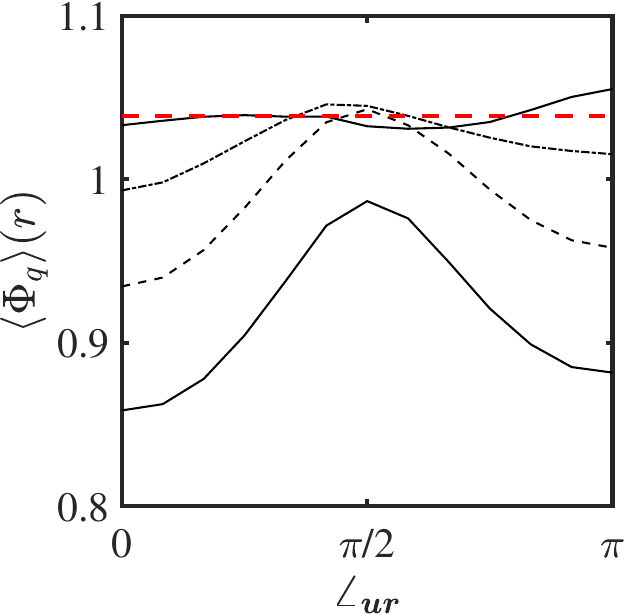}}%Phivsang_f10}}%
\mylab{-.16\textwidth}{.32\textwidth}{(b)}%
\hspace*{5mm}%
\raisebox{0mm}{\includegraphics[height=0.30\textwidth,clip]{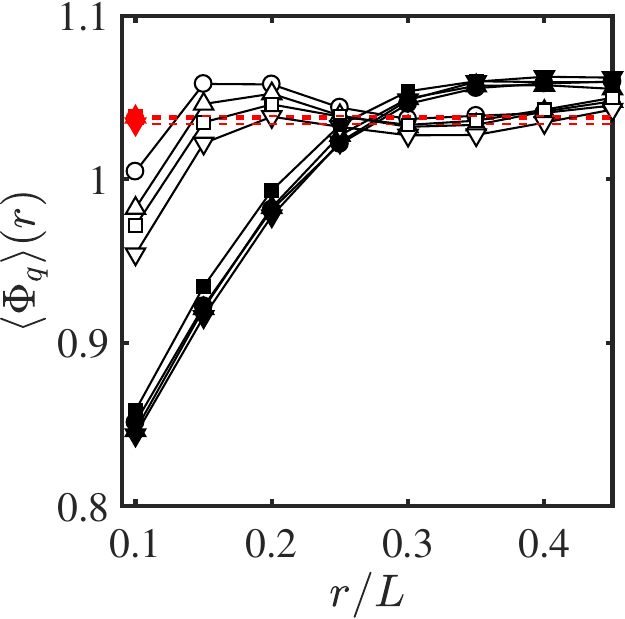}}%Phimaxmin_f10}}%
\mylab{-.16\textwidth}{.31\textwidth}{(c)}%
}%
\caption{%
Mean approximation error conditioned to the orientation with respect to the velocity.
(a) Definition sketch.
(b) Mean error as a function of the orientation angle. T1024 at $\omega'_0 t=39$. From bottom to top, 
distance to the reference point: $r/L=0.1\, (0.05)\, 0.25$. 
(c) Streamwise and transverse conditional errors as functions of the distance to the reference
point. Symbols as in table \ref{tab:cases}, with each case at its final simulation time.
Closed symbols are measured aligned with the velocity, and open ones are measured at right angles to
it. The dashed horizontal lines in (b,c) are unconditional errors.
}
\label{fig:angleu}
\end{figure}
% ===========================================================

Figure \ref{fig:angleu} tests whether the structures detected by the concatenation of
dipoles are elongated and aligned with the flow velocity, as suggested by figure
\ref{fig:heat}(c). Consider the sketch in figure \ref{fig:angleu}(a). For a given flow
snapshot and template size, each cell in the $N_t\times N_t$ test grid has an associated
flow velocity, $\vec{u}$ and an approximation error, $\Phi$, defined by averaging over the
domain of the optimal template centred on it. Choosing a displacement vector $\vec{r}$ with
respect to this point, figure \ref{fig:angleu}(b,c) shows the average of the approximation
error corresponding to the cell whose centre is closest to the end of $\vec{r}$. The figure
shows the mean error conditioned to the magnitude of $\vec{r}$ and to its angle,
\mbox{\boldmath$\angle_{ur}$}, with respect to $\vec{u}$. When this conditional mean is
averaged over all the points of the test grid, the result is essentially independent of the
angle, and similar to the unconditional mean of the error. But, when the conditional centre
is chosen within the largest low-error structure in each flow field, figure
\ref{fig:angleu}(b) shows that points aligned with the velocity,
$\mbox{\boldmath$\angle_{ur}$}=0$ or $\mbox{\boldmath$\angle_{ur}$}=\pi$, preferentially
contain low approximation errors, while those perpendicular to it have high ones. This
effect weakens with the length of $r$, but figure \ref{fig:angleu}(c) shows that it persists
for a distance of the order of $0.25 L$, which is four or five times longer than the width
$\rho_1$ in figure \ref{fig:heath}(a). It is interesting that the distance at which the
minimum error in figure \ref{fig:angleu}(c) reverts to its unconditional value is
approximately the same for the four cases included in the figure. In fact, it changes little
among all the cases tested, showing that the longitudinal scale of the high-velocity streams
is always of the order of the box size. The growth of the energy wavelength, $\lambda_q$, in
figure \ref{fig:specs}(b) and of the area of the high-energy region in figure
\ref{fig:heath}(a) is presumably due to more convoluted streams, rather than to longer-range
ones.

An intriguing feature of figure \ref{fig:angleu}(b) is the asymmetry between the conditional
error at $\mbox{\boldmath$\angle_{ur}$}=0$ and $\mbox{\boldmath$\angle_{ur}$}=\pi$. Since
each point in the low-error structures is both the origin and the end of some conditioning
vector, both directions could be expected to be equivalent. But they are not, and the error
is lower in front of the conditioning point than behind it. The effect is small but
statistically significant; the estimated standard deviation of the lines in figure
\ref{fig:angleu}(b) is only slightly larger than the width of the lines. It is also
consistently found in all the cases examined, not only in the one chosen for the figure.
This result is difficult to interpret, but the implication is that the direction of the velocity is a better
predictor of the downstream location of the low-error structures than of their upstream
location. We will come back to this point in \S\ref{sec:tur2d}.

%----------------------------------------------------------------------------------------------------------------
\subsection{Vortices}\la{sec:vortices}

Even if this paper is mostly concerned with the large energy-containing structures of the
flow, there is no doubt that two-dimensional turbulence can also be described as a
collection of coherent vortices \citep{mcwilliams84,mcwilliams90,Benzi92}. The question that
interests us here is whether the large-scale structure discussed above for the kinetic
energy can be described in terms of the organisation of these vortices. In this section, we
first address the properties and evolution of the vortices themselves.

% ===========================================================
\begin{figure}
\vspace*{6mm}%
\centerline{%
\raisebox{7.5mm}{\includegraphics[height=0.253\textwidth,clip]{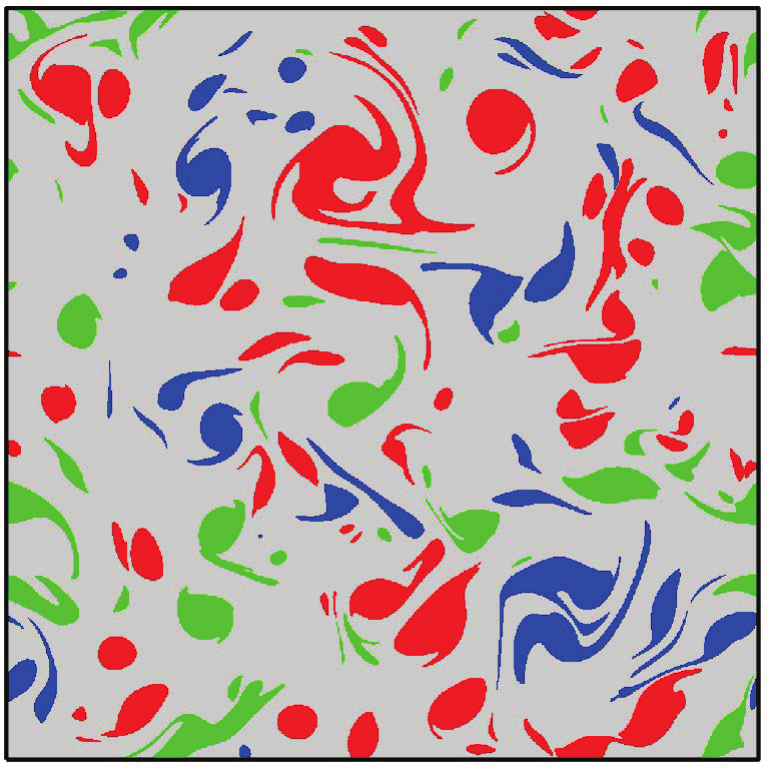}}%pairs.png}}%
\mylab{-.15\textwidth}{.32\textwidth}{(a)}%
\hspace*{3mm}%
\raisebox{0mm}{\includegraphics[height=0.31\textwidth,clip]{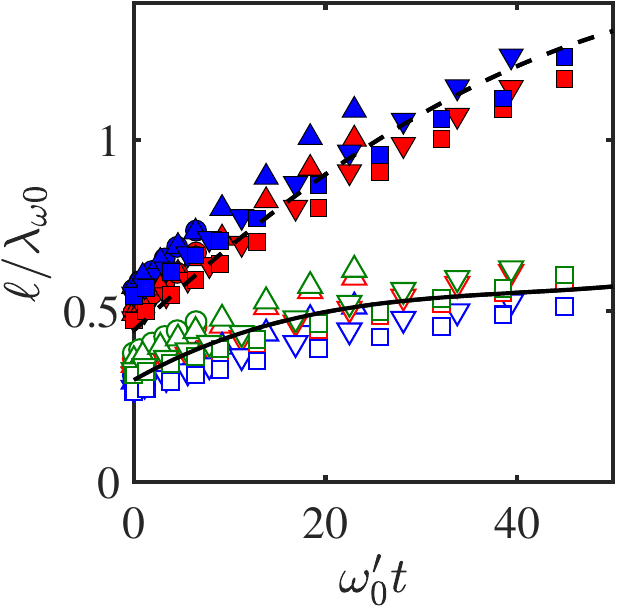}}% vortexsize_mism2}}%
\mylab{-.15\textwidth}{.32\textwidth}{(b)}%
\hspace*{3mm}%
\raisebox{0mm}{\includegraphics[height=0.31\textwidth,clip]{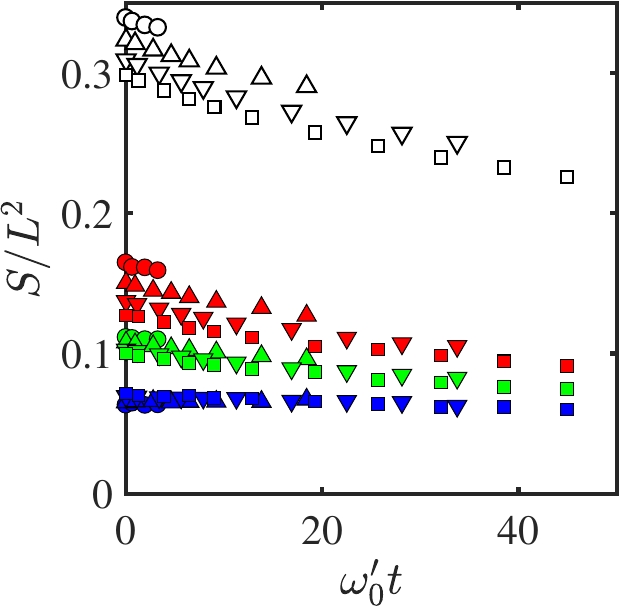}}%  areafrac}}%
\mylab{-.15\textwidth}{.32\textwidth}{(c)}%
}%
\vspace{5mm}%
\centerline{%
\raisebox{0mm}{\includegraphics[height=0.31\textwidth,clip]{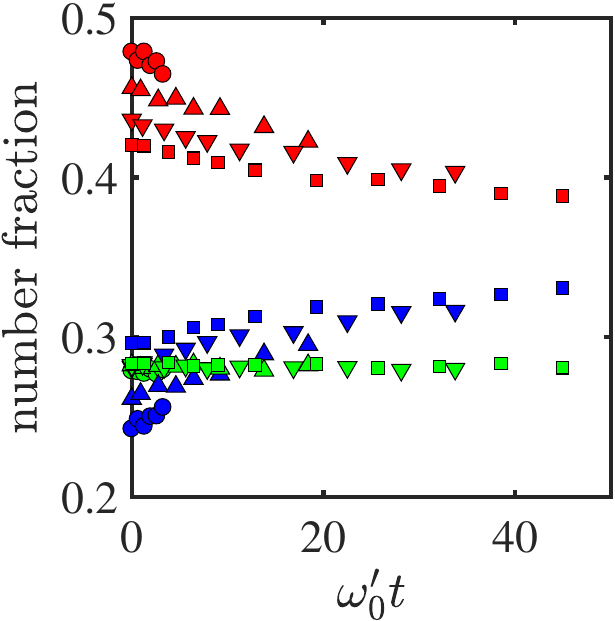}}%  vortexnum_mism2}}%
\mylab{-.15\textwidth}{.32\textwidth}{(d)}%
\hspace{3mm}%
\raisebox{0mm}{\includegraphics[height=0.305\textwidth,clip]{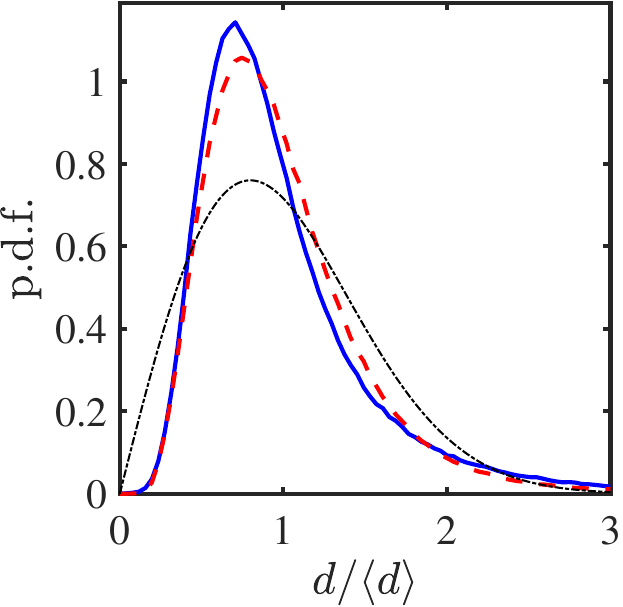}}% distanceall}}%
\mylab{-.15\textwidth}{.32\textwidth}{(e)}%
\hspace{3mm}%
\raisebox{0mm}{\includegraphics[height=0.31\textwidth,clip]{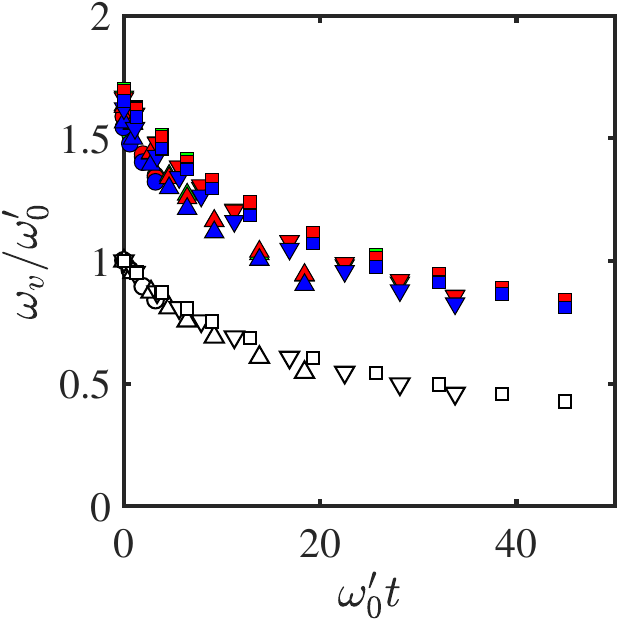}}% corevorticity}}%
\mylab{-.15\textwidth}{.32\textwidth}{(f)}%
}%
\vspace{5mm}%
\centerline{%
\raisebox{0mm}{\includegraphics[height=0.31\textwidth,clip]{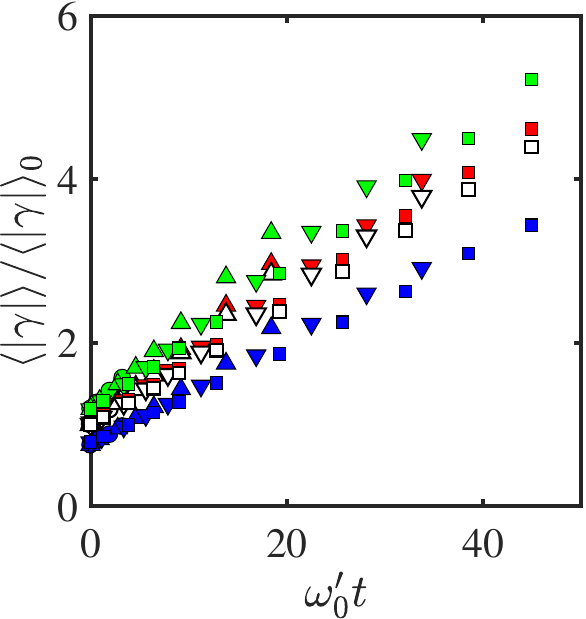}}%  circulation}}%
\mylab{-.15\textwidth}{.32\textwidth}{(g)}%
\hspace{3mm}%
\raisebox{0mm}{\includegraphics[height=0.31\textwidth,clip]{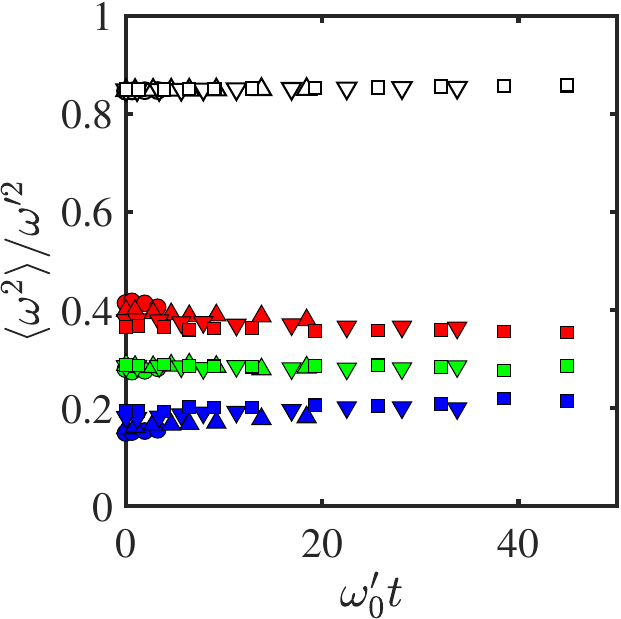}}%  enstrophyfrac}}%
\mylab{-.15\textwidth}{.32\textwidth}{(h)}%
\hspace{3mm}%
\raisebox{0mm}{\includegraphics[height=0.31\textwidth,clip]{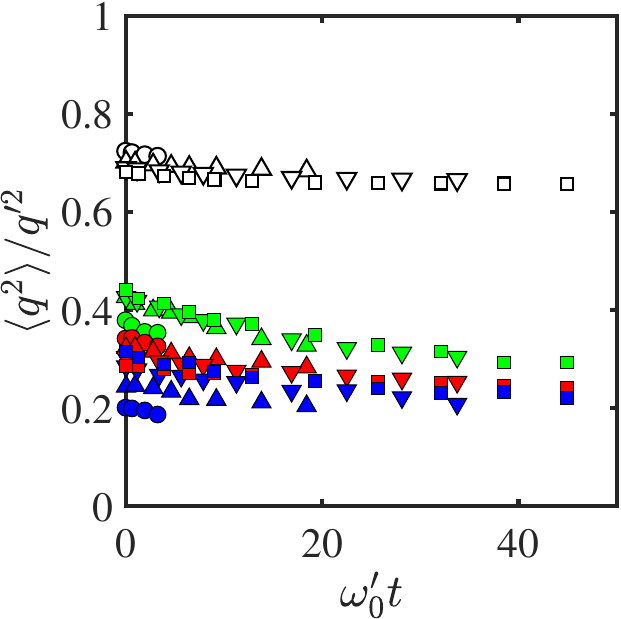}}% energyfrac}}%
\mylab{-.15\textwidth}{.32\textwidth}{(i)}%
}
\caption{%
Properties of the thresholded vorticity structures. 
(a) Typical segmented image. Case T1024, $\omega'_0t=39$. In all the panels in this figure, unless otherwise
noted, red are dipoles, blue are corrotating pairs, and green are isolated vortices. Other
symbols as in table \ref{tab:cases}.
(b) The open symbols are the mean
diameter of the vortices, and the closed coloured ones are the distance between vortices in a pair.
Lengths are normalised with the vorticity wavelength at $t=0$, and the trend lines are the
polynomial fits from figure \ref{fig:specs}(b) to: \solid, $0.3\lambda_\omega$; \dashed, $0.22 \lambda_q$.
(c) Area fraction covered by all the vortices of a given class. Open symbols include all the classes.
(d) Number fraction of vortices in different  associations. 
(e) P.d.f. of the distances between vortices in a pair. Colours as in (a), but the chain-dotted black line is the
distribution for a Poisson point set.
(f) Mean vorticity of the vortex cores. Symbols in (f--i) are as in (c).
(g)  Average circulation of the vortex cores. 
(h) Enstrophy fraction contained in the different  vortex associations.  
(i) As in (h), for the kinetic energy.
}%
\label{fig:vortices}
\end{figure}
% ===========================================================

Figure \ref{fig:vortices}(a) shows a segmentation of a typical flow field into individual
vortices, defined as connected regions in which $|\omega|\ge H\omega'$. As in the case of
figure \ref{fig:heath}, the vorticity of the flow separates into a few large connected
objects for $H\ll 1$, and breaks into more numerous smaller objects as $H$ increases. Beyond
a certain threshold, the number of vortices decreases again, and eventually vanishes when no
vorticity satisfies the thresholding condition. The value $H=0.9$ used in figure
\ref{fig:vortices} is chosen to maximise the number of individual vortices \citep{moisy04}.
To gain some sense of the importance of vortex interactions, the vortices in figure
\ref{fig:vortices}(a) are grouped into co- and counter-rotating pairs. Two vortices are
considered a potential pair if their area, $s$, differs by less than a factor of $m^2$,
which is an adjustable parameter. The underlying rationale for this restriction is
that very dissimilar vortices are unlikely to form long-lived pairs, because the larger one
would tear the smaller one apart \citep{meunier05}. The figure uses $m=2$, but statistics
compiled with $m=1.5$ and $m=3$ show no substantial differences \citep[see][]{jotploff}.
Vortices are paired to the closest unpaired neighbour within their area class, and no vortex
can have more than one partner. Some vortices find no suitable partner, and are left
unpaired.

Figure \ref{fig:vortices}(b) displays mean values of the diameter, $s^{1/2}$, and of the
distance $d$ between the centres of gravity of the component vortices of the pairs, compiled
at several evolution times for each set of simulations. The diameter of the vortices depends
very little on how they are paired, and, in agreement with the estimations from the
vorticity and velocity statistics in figure \ref{fig:specs}, is described well by a fraction
of the vorticity wavelength, $0.3 \lambda_\omega$. The distance between the vortices of a
pair is initially somewhat larger, of the order of $0.5 \lambda_\omega$, but it grows with
$\lambda_q$, faster than $\lambda_\omega$. The approximate similarity between the
inter-vortex distance and their diameter implies that vortex pairs remain tightly packed at
this stage of their evolution, although the faster growth of the average distance means that
the area fraction covered by the vortices slowly decays (figure \ref{fig:vortices}c). It
should be noted at this point that the absolute value of the intra-pair distance in figure
\ref{fig:vortices}(b) offers a possible explanation of the observation in \S\ref{sec:large}
that the edges of the high-speed streams are not a concentration of high enstrophy. It
follows from comparing figures \ref{fig:heath}(a) and \ref{fig:vortices}(b) that $d$ is 2--3
times narrower than the width $\rho_1$ of the jets, so that, if these jets are defined as in
figure \ref{fig:heath}, the dipoles are contained within them, rather than at their border.
This is consistent with the visual inspection of the snapshot in figure \ref{fig:heat}(b).

Figure \ref{fig:vortices}(d) shows the number of vortices involved in different kinds of
pairings, giving a rough measure of the importance of the different interactions. Most
vortices are in the form of pairs. Of the approximately $2\times 10^6$ vortices represented
in figure \ref{fig:vortices}, 43\% form dipoles, 29\% are in corrotating pairs, and 28\% are
isolated, with a tendency of the number of dipoles and corrotating pairs to converge as the
Reynolds number increases. Similar values were found by \cite{jotploff} at the lowest of the
four Reynolds numbers used in the present paper. The difference between the number of
corrotating and counterrotating pairs is also encountered in the covering fraction in figure
\ref{fig:vortices}(c), and is a property of the flow that disappears if the vortex position
is randomised. On the other hand, the scarcity of unpaired vortices is a geometric property
that persists when the pairing algorithm is applied to a set of Poisson-distributed points.

In fact, a random distribution of the vortex position is a reasonable lowest-order model for
their local organisation. Even if the vortex diameter and the intra-pair distance grow by a
factor of approximately two during the simulation time, the form of their p.d.f. stays
remarkably constant. For example, figure \ref{fig:vortices}(e) shows the p.d.f. of the
distance among the components of vortex pairs, normalised by their mean for each individual
experiment. The figure also includes the p.d.f of a set of Poisson-distributed points with
the same mean. The distribution of dipoles and corrotating pairs are very similar, with a
weak tendency of dipoles to be farther apart. The Poisson distribution is wider, but most of
the discrepancy can be explained by the exclusion of pairs whose separation is smaller than
the vortex diameter (see appendix \ref{sec:poisson}). This explanation is confirmed by the
distribution of the distance $d$ between all possible couples of vortices (not shown). This
distribution should be proportional to $d$ for any homogeneous flow, but this is only true
for $d \gtrsim 2\bra s^{1/2}\ket$. Shorter distances are essentially missing.

In the same way, there is relatively little difference between the properties of vortices in
dipoles and those in corrotating pairs. We saw in figure \ref{fig:vortices}(b) that their
diameters are similar, and figure \ref{fig:vortices}(f) shows that so are their vorticities,
which decay at the same rate as the r.m.s. vorticity of the flow. On the other hand, some
explanation is needed for figure \ref{fig:vortices}(c,d), which shows that both the number
of dipoles and the total area covered by them is larger than those of corrotating pairs. A
similar difference appears in figure \ref{fig:vortices}(g), which shows that the average
vortex circulation magnitude grows linearly for all classes. The circulation of a vortex core can only
grow by merging with other cores or by entraining background vorticity, since there is
no vorticity source in two-dimensions, but the average over a class is also influenced by the
transfer among classes. Figure \ref{fig:vortices}(g) shows that the circulation of dipoles
is typical of the overall average (they are the largest contributors to it), but that
corrotating pairs grow more slowly, while unpaired vortices grow slightly faster than the
average. The simplest explanation is that corrotating pairs tend to merge into single
(initially unpaired) cores \citep{meunier05}, thus depleting their number, while dipoles are
longer-lasting \citep{Flierl:80,mcwilliams80}. 

Figure \ref{fig:vortices}(h) shows the fraction of overall enstrophy contained in the
thresholded vortices. Over 85\% of the total enstrophy is contained in them, and this
fraction is remarkably constant among Reynolds numbers and time, no doubt, in part, because
the vorticity threshold used to identify them is a constant fraction of the r.m.s.
vorticity. The contribution to the enstrophy of the different vortex classes is in line with the area
fraction in figure \ref{fig:vortices}(c), as could be expected from the similarity of vortex
properties discussed above.

The fraction of the kinetic energy due to the vortices is harder to define. The simplest
definition is the fraction of $q'^2$ retained by a flow reconstructed from the vorticity contained
within a given class of thresholded vortices. Figure \ref{fig:vortices}(i) shows that
keeping all the thresholded vortices retains 65--70\% of $q'^2$, while keeping only the dipoles or
the corrotating pairs retains 25-30\% of the energy, and keeping only the unpaired
vortices retains 30--40\%. The change in the relative contribution of dipoles to
the enstrophy in figure \ref{fig:vortices}(h) and to the energy in figure
\ref{fig:vortices}(i) is interesting. While dipoles predominate in number, area, and
enstrophy, their contribution to the energy is of the same order as the corrotating pairs,
and substantially less than the unpaired vortices. In fact, even if dipoles contain a local
jet of high velocity, their overall kinetic energy is lower than for corrotating pairs, because the
total circulation of a dipole is zero, and its induced velocity falls with distance much
faster than for a corrotating pair \citep{bat67}. Even if decaying turbulence is very far
from an equilibrium system, \cite{Benzi92} and \cite{dritsch08} have shown that an approximately Hamiltonian
system of point vortices, punctuated by the occasional merger of like-signed vortices, is a
good approximation to the late stages of two-dimensional enstrophy decay. The difference in the
interaction energy of dipoles and corrotating pairs is probably also part of the reason for their
different behaviour, and for the slower decay of the former through amalgamation.
 
In summary, the vortex evolution discussed in this section is consistent with the description of
decaying two-dimensional turbulence as a system of discrete vorticity structures
\citep{mcwilliams84,Benzi87,brach88}, although, at the relatively early stage of the decay
studied here, the structures cannot be described as either equilibrium or isolated. Their
self-similar growth through amalgamation accounts for the gradual increase in the vorticity
scale, $\lambda_\omega$, but the sizes involved are always much smaller than the energy
scale, $\lambda_q$ (see the scale bars in figure \ref{fig:omfields}).

It is interesting that the Reynolds number of the individual vortex cores stays
relatively low during the simulations, $Re_\gamma=\bra\gamma\ket /4\pi\nu \approx 1.5
\to 7$. Since the maximum vorticity of a viscously decaying circular vortex satisfies
$\omega t =Re_\gamma$ \citep{bat67}, where the left-hand side is time measured in turnovers,
it follows that the size of the vortices in the simulations, even as they grow by accretion,
is typically controlled by viscosity.

% ---------------------------------------------------------------------------------------------------------
\section{Collective structures}\la{sec:collective}

% -------------------------------------------------------------------------------------------------------------
\subsection{Vortex organisation}\la{sec:crystals}

In fact, the question of how vortices organise themselves to create the streams
discussed in \S\ref{sec:large} remains open. One possibility, already mentioned, is that the
streams are concatenations of dipoles that are responsible for short segments of the stream. We
saw in the discussion of figure \ref{fig:vortices} that this model is compatible with the
observed vortex distances and dimensions, but the open question is how the individual
dipoles align themselves into longer units. Another model is that the streams are contact
interfaces between large-scale vortices. We saw in figure \ref{fig:vortices}(b) that the
mean vortex diameter increases only weakly during even the longest flow evolution, and that
it is always much smaller than the kinetic-energy wavelength, but it is possible that the
small cores organise themselves into large-scale vortex `bags' that fill the space between
the streams, acting as coherent structures from the point of view of the kinetic energy
\citep{Par:Tab:98,tabeling02}. However, we saw in \S\ref{sec:large} that the only
concentration of vorticity is within the coherent jets, not between them.

\begin{figure}
\vspace*{5mm}%
\centerline{%
\raisebox{0mm}{\includegraphics[height=0.30\textwidth,clip]{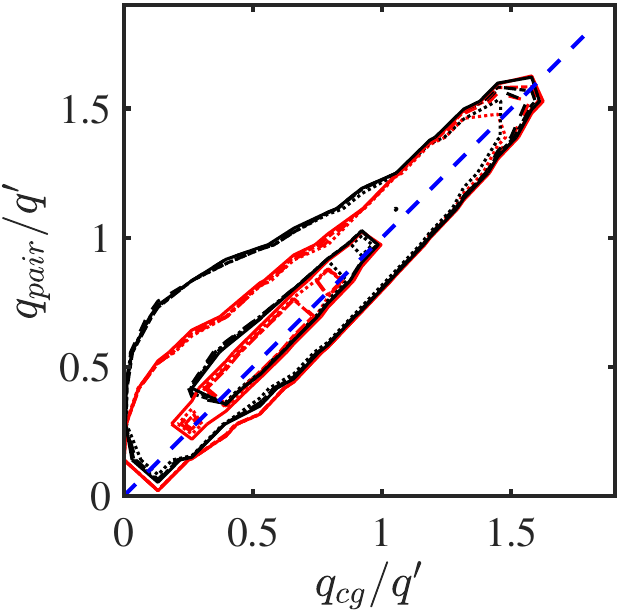}}%vcg_vvor}}%
\mylab{-.16\textwidth}{.32\textwidth}{(a)}%
\hspace*{5mm}%
\raisebox{0mm}{\includegraphics[height=0.305\textwidth,clip]{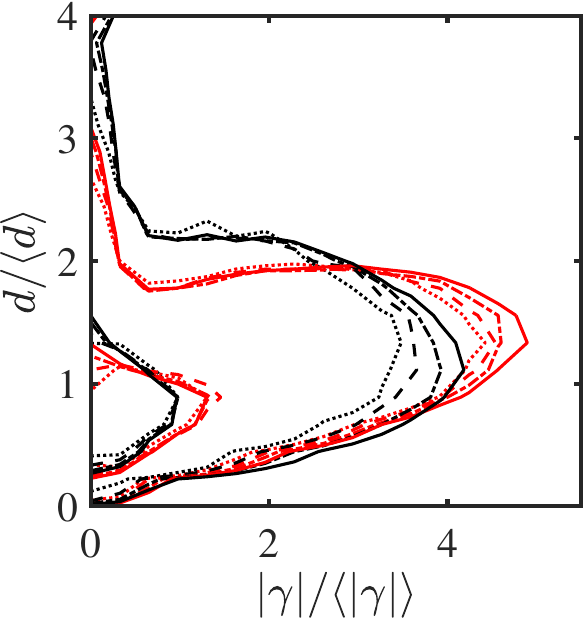}}%circ_dist}}%
\mylab{-.16\textwidth}{.32\textwidth}{(b)}%
\hspace*{5mm}%
\raisebox{0mm}{\includegraphics[height=0.30\textwidth,clip]{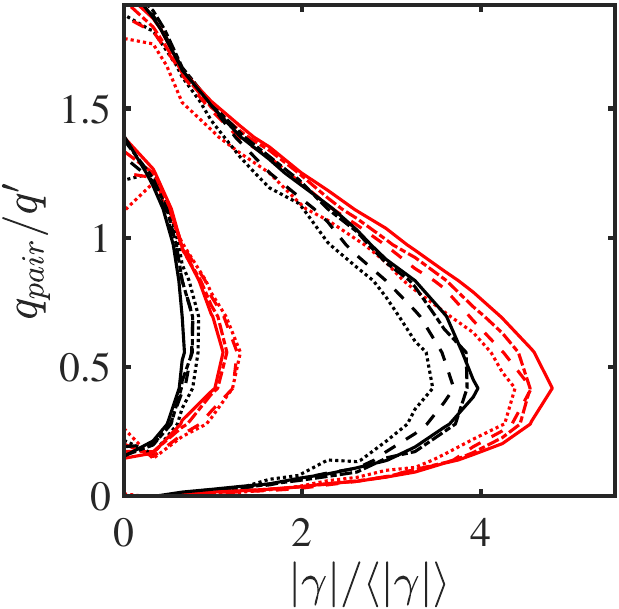}}%circ_vvor}}%
\mylab{-.16\textwidth}{.31\textwidth}{(c)}%
}%
\caption{%
Properties of the vortex pairs.
(a)  Joint p.d.f. of the velocity of the centre of gravity of a vortex pair versus the averaged velocity
magnitude of its two component vortices. In all the panels in this figure: \dotted, T256;
\dashed, T512; \chndot, T768; \solid, T1024. Red lines are dipoles, and black ones are
corrotating pairs. The two probability contours in each case enclose 50\% and 95\% of the probability mass. 
(b) Mean circulation magnitude of the vortex components of the pair, versus the 
inter-component distance.
(c) Vortex circulation versus vortex mobility.
}
\label{fig:selfvort}
\end{figure}

The crucial uncertainty is the intensity of the interaction among vortex cores, and whether,
for example, the advection velocity of the cores is mainly due to their closest neighbour, or to
a background of `field' vortices. The former would support the first of the two
models above, while the latter would support the second. Consider vortex pairs. The result of the mutual
induction among two vortices of the same sign and similar circulation is a rotation around
each other. If the vortices are denoted by $A$ and $B$, and we estimate the (vector) `mobility' of a vortex
by averaging the flow velocity over its core,
\beq
\vec{u}_{vor}=s_{vor}^{-1}\int_{vor} \vec{u} \dd s,
\la{eq:uadv}
\eeq
the induced mobilities of the components of a corrotating pair would be $\vec{u}_A= -\vec{u}_B$, and the
velocity of the centre of the pair would vanish, $\vec{u}_{cg}=(\vec{u}_A+\vec{u}_B)/2=0$.
On the other hand, dipoles self-induce a common translation velocity, and
$\vec{u}_A=\vec{u}_B=\vec{u}_{cg}$. If we define $q_{vor}= \|\vec{u}_{vor}\|$, a self-inducing 
corrotating pair would be characterised by $q_{cg}\ll q_{pair}\equiv (q_A+q_B)/2$,
while a dipole would satisfy $q_{cg}\approx q_{pair}$. This is tested in figure
\ref{fig:selfvort}(a), and it is only satisfied by slow-moving vortices. The behaviour of
fast pairs for which $q\gtrsim q'$ is independent of whether they are co- or
counter-rotating, and they can therefore be assumed to be mostly advected by a background
velocity field.

% ===========================================================
% =========================================================== 

Figure \ref{fig:selfvort}(b) shows that vortex pairs can be classified into two groups. The
`nose' extending to the lower right in the figure represents a family of strong cores with
large circulations, whose intra-pair distance is relatively small. This family exists for
corrotating pairs and for dipoles, although it is most marked for the latter. The vertical
band to the left of the figure contains relatively weak vortices with no clear preference
for a particular coupling distance. Most vortex pairs are in this latter family, but they are
relatively unimportant for the flow. Approximately 66\% of the cores have
$|\gamma|<\bra|\gamma|\ket$, but they only contain 15-25\% of the total circulation magnitude. A
similar distinction can be based on vortex area (not shown), since the mean vorticity is
relatively uniform among the vortices. The classification into vortex types can be
based on either property.

One could hypothesise that the large, tightly-coupled vortices in the `nose' family would be
the ones with the fastest mobilities, which they would induce on each other, but figure
\ref{fig:selfvort}(c) shows that the opposite is true. The vortices with the largest circulations move
relatively slowly, and the high mobilities tend to be associated with the weak
circulations to the left of the figure. This somewhat surprising observation leads  to a model
in which a set of organising large vortices in an `approximate equilibrium' configuration
are responsible for organising the flow into streams where weaker
vortices are advected at relatively high speed.

% ===========================================================
\begin{figure}
\vspace*{5mm}%
\centerline{%
\raisebox{0mm}{\includegraphics[height=0.307\textwidth,clip]{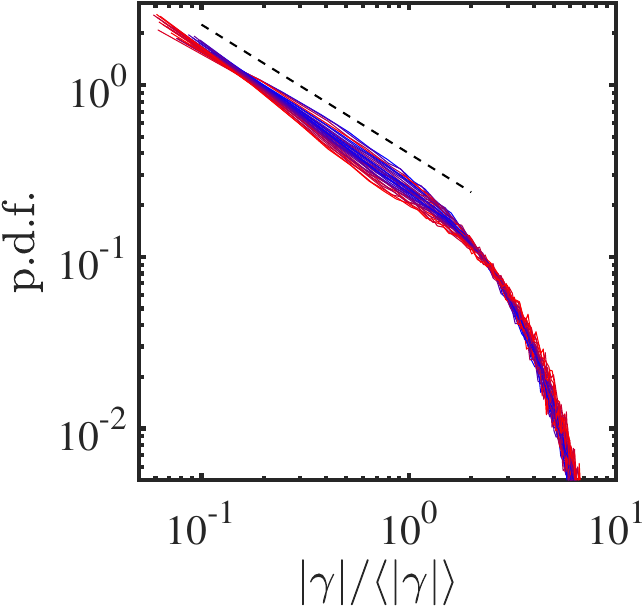}}%gammaloglog}}%
\mylab{-.15\textwidth}{.32\textwidth}{(a)}%
\hspace*{2mm}%
\raisebox{0mm}{\includegraphics[height=0.305\textwidth,clip]{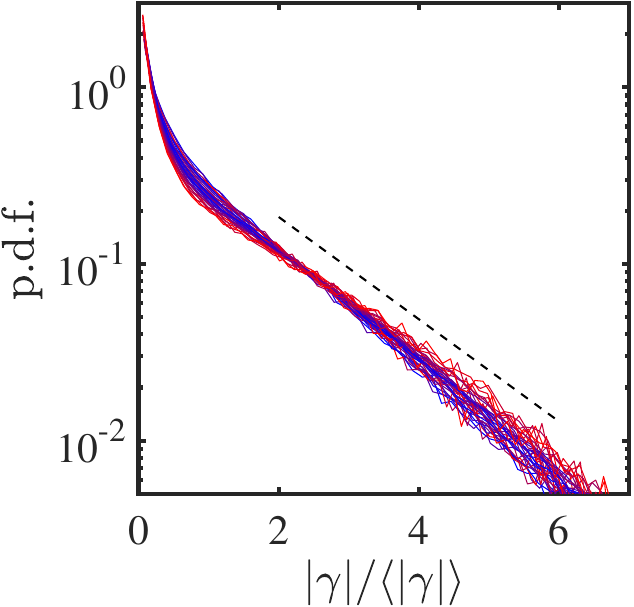}}%gammalogy}}%
\mylab{-.15\textwidth}{.32\textwidth}{(b)}%
\hspace*{2mm}%
\raisebox{0mm}{\includegraphics[height=0.31\textwidth,clip]{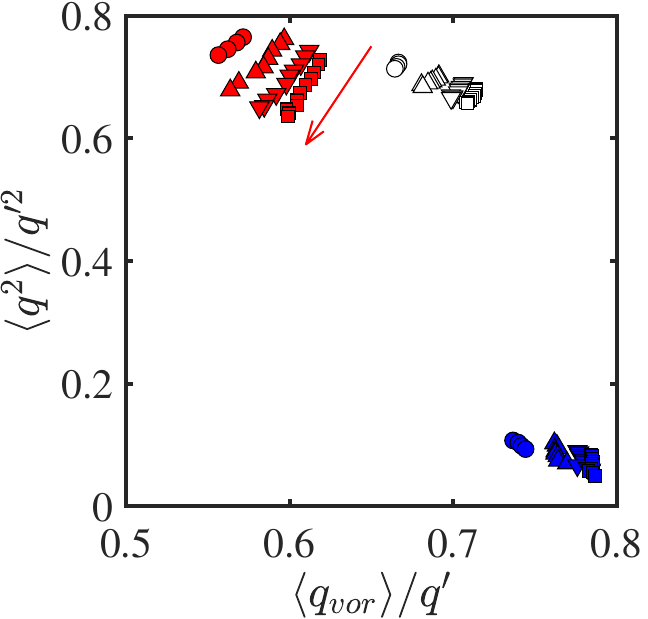}}%energyadvecL}}%
\mylab{-.15\textwidth}{.32\textwidth}{(c)}%
}%
\caption{%
(a) P.d.f. of the vortex circulation magnitude. In all cases, time increases from blue to red. The
dashed line is proportional to $|\gamma|^{ -0.75}$.
(b)  As in (a). The dashed line is proportional to $\exp(-0.666 |\gamma|/ \bra |\gamma|\ket)$.
(c) Kinetic energy contained in different vortex classes versus mean vortex mobility.
Open symbols are all the thresholded vortices; red
symbols are large vortices, $s_{vor} \ge \bra s_{vor}\ket$; blue symbols are $s_{vor} < \bra s_{vor}\ket$.
The red arrow is the direction of time advance for the red symbols.
}
\label{fig:families}
\end{figure}
% ===========================================================

In fact, the large and small vortices have very different properties. Figure
\ref{fig:families}(a,b) shows the one-dimensional p.d.f. of the core circulation, which
collapses reasonably well for all the simulations and evolution times. Figure
\ref{fig:families}(a) shows that the weak vortices, $|\gamma|< \bra |\gamma|\ket$, follow a
power-law distribution $P(|\gamma|)\sim |\gamma|^{-3/4}$, while figure \ref{fig:families}(b)
shows that the vortices above that limit follow an exponential one. The separation of
decaying two-dimensional turbulence into coherent vortices evolving under mutual induction,
and a `chaotic' background has been discussed often \citep{Benzi87,Benzi88,mcwilliams90b},
but the background is usually not characterised in term of vortices, and we are not aware of
any previous characterisation of the larger vortices as slowly moving. For example,
\cite{Benzi87} discuss the large vortices as the only coherent structures in the flow and
report that their areas follow a power-law distribution, although with a different exponent
than the one above, $|\gamma| ^{-3/2}$. The reason for this difference is not clear,
but we mentioned in \S\ref{sec:simul} that \cite{Benzi87} study a later stage of the
decay, and that they use a higher-order viscous dissipation, which presumably creates
different vortex cores \citep{mcwilliams90b,jim:hyper:94}. It is also unclear why the
exponential range of the distribution in figure \ref{fig:families}(b) is not discussed by
\cite{Benzi87}, although this may be partly due to their different sample size. The analysis
in their paper is based on 17 vortices, while each of the distributions in figure
\ref{fig:families} represents $10^4$--$10^5$ objects.

Figure \ref{fig:families}(c) shows that the kinematics of the weak and strong cores is 
very different. The vertical axis in this figure is the fraction of the kinetic energy carried
by the thresholded vortices, defined as in figure \ref{fig:vortices}(i). The open symbols
are the contribution from all the vortices, as in that figure, and the red symbols are the
contribution from vortices whose area is larger than the average. They contain most of the
kinetic energy. The blue symbols in the lower-right corner are the contribution from 
vortices smaller than the average, which is much smaller. The horizontal axis is the average
vortex mobility defined as the modulus of \r{eq:uadv}. The large and small vortices lie in very
different parts of the plot, as already suggested by figure \ref{fig:selfvort}(c). Large vortices
are responsible for most of the kinetic energy of the flow, but are themselves relatively
immobile, while small ones move fast, but are only responsible for a small fraction of the
kinetic energy.

The power-law and exponential probability distributions suggest that the cores grow by two different
aggregation mechanisms of smaller units. While a power law implies self-similar scale-free growth, in
which cores merge with other cores of similar size \citep{Benzi92}, an exponential has a
definite scale, which is proportional to the mean of the distribution, and to the size of
the elements being accreted \citep{JimKaw13}. In figure \ref{fig:families} the lower limit
of the exponential is indeed of the order of $\bra|\gamma|\ket$, and it is interesting to
speculate about an aggregation model in which cores merge self-similarly with each other until
they grow to be large enough to `freeze' in a quasi-equilibrium slowly evolving
pattern. The motion of these large vortices is not chaotic, at least over short times, and
these cores stop merging among themselves. But they keep absorbing the remaining fast-moving
vortices of the background, and the largest of these field vortices determine the scale of the
exponential distribution.

Over a much longer time scale, it is to be expected that even the `frozen' vortices would
merge among themselves, in an amalgamation process similar to the one described by
\cite{carnevale91,carnevale92}, but the appearance of a collective equilibrium signals that
the energy scale has become of the order of the box size. As mentioned in \S\ref{sec:simul},
this is the limit of our simulations, and of the analysis in this paper.

Vortex arrangements that remain stationary in some frame of reference have been studied for
over a century \citep[see the review in][]{aref02}. Some of them are stable, and form
spontaneously in experiments. In particular, forced two-dimensional turbulence is known to
settle to stationary vortex `crystals' which are partly determined by the forcing method and
by the boundary conditions \citep{fine95,jin00b,jim:gueg:07}, and beautiful examples of
equilibrium vortex polygons have been observed in the polar regions of planetary
atmospheres \citep{taba:etal:20}. Most known equilibrium systems are regular arrangements of
vortices of a single sign in a background of opposite-sign vorticity, but mixed-sign stable
systems are also known. The von K\'arm\'an vortex street is probably the best-known example
of the latter, and it is known that two-dimensional turbulence in a square box converges to
a quasi-equilibrium single dipole, arranged diagonally in the box, which only decays slowly
by viscosity \citep{smithyak93}.

The circulation of the  slow-moving vortices discussed here is essentially in balance
(with a residue of 1\%-3\% of $\omega'$), but they are still far from equilibrium, and may
perhaps be considered an intermediate stage to a final steady state. Attempts to extract a
regular arrangement for them failed, beyond the four-way symmetry induced by the computational box,
but the drop in mobility can be considered diagnostic of incipient `crystallisation'.

% ===========================================================
\begin{figure}
\vspace*{4mm}%
\centerline{%
\raisebox{0mm}{\includegraphics[height=0.23\textwidth,clip]{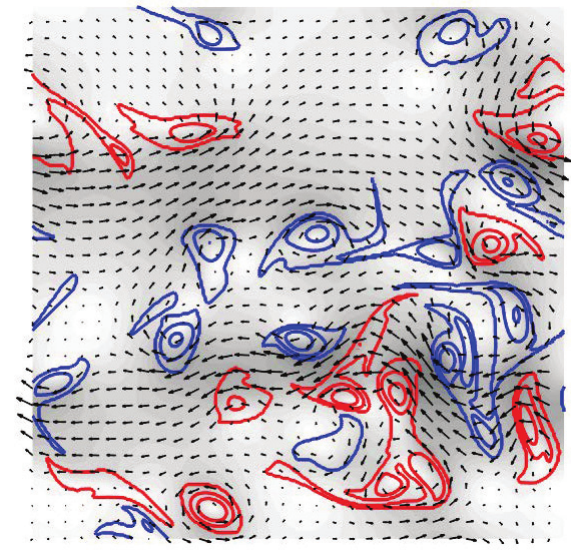}}% noploff256_t10_test9_tim0}}%
\mylab{-.27\textwidth}{.11\textwidth}{(a)}%
\hspace*{2mm}%
\raisebox{0mm}{\includegraphics[height=0.23\textwidth,clip]{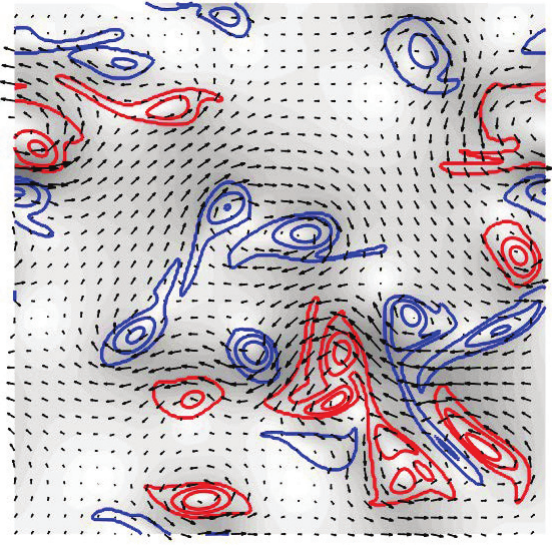}}% noploff256_t10_test9_tim3}}%
\hspace*{2mm}%
\raisebox{0mm}{\includegraphics[height=0.23\textwidth,clip]{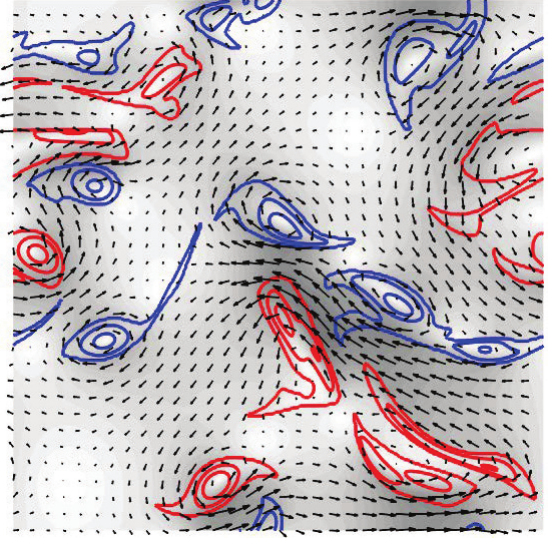}}% noploff256_t10_test9_tim7}}%
\hspace*{2mm}%
\raisebox{0mm}{\includegraphics[height=0.23\textwidth,clip]{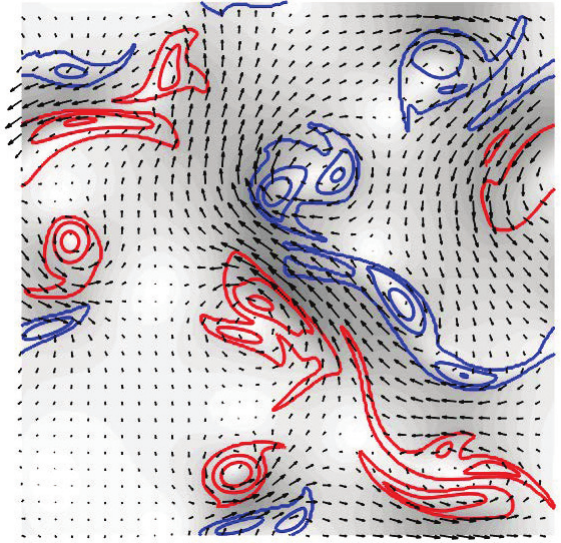}}% noploff256_t10_test9_tim10}}%
}%
\vspace*{3mm}%%%%%%%%%%%%%%%%%%%%%%%%%%%%%%%%%%
\centerline{%
\raisebox{0mm}{\includegraphics[height=0.23\textwidth,clip]{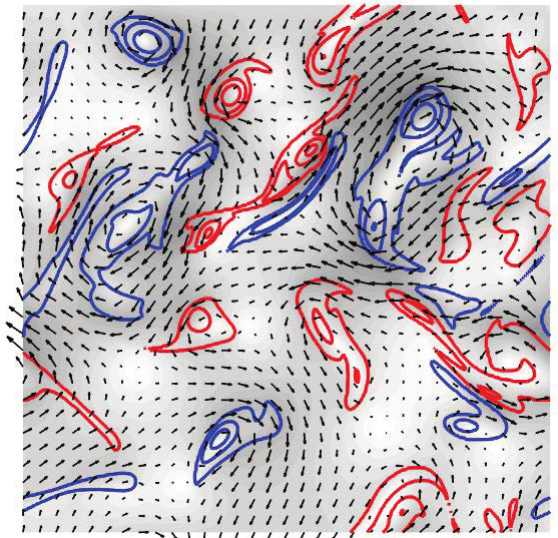}}% noploff256_t10_test7_tim0}}%
\mylab{-.27\textwidth}{.11\textwidth}{(b)}%
\hspace*{2mm}%
\raisebox{0mm}{\includegraphics[height=0.23\textwidth,clip]{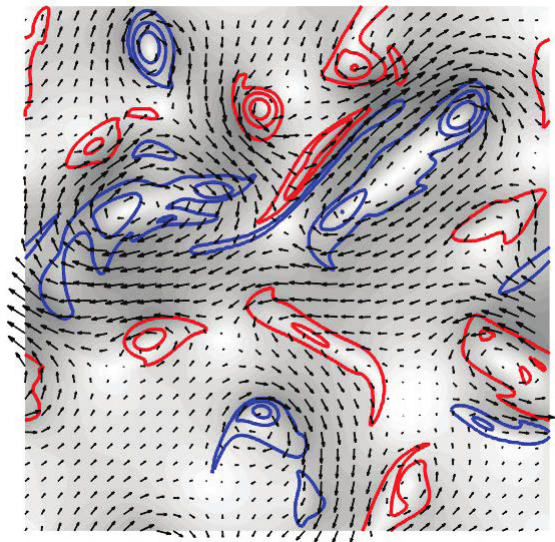}}% noploff256_t10_test7_tim3}}%
\hspace*{2mm}%
\raisebox{0mm}{\includegraphics[height=0.23\textwidth,clip]{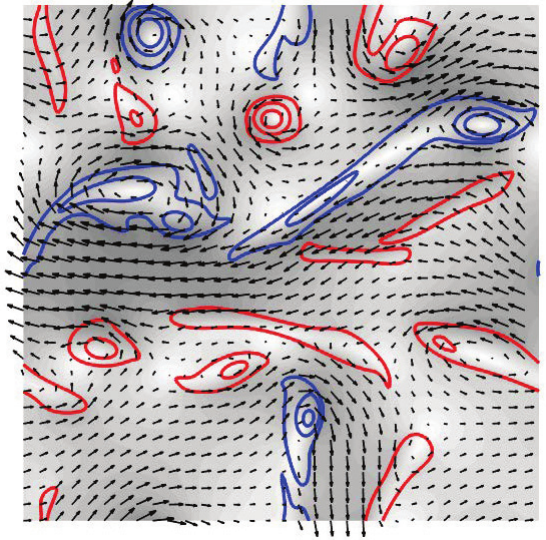}}% noploff256_t10_test7_tim7}}%
\hspace*{2.5mm}%
\raisebox{0mm}{\includegraphics[height=0.23\textwidth,clip]{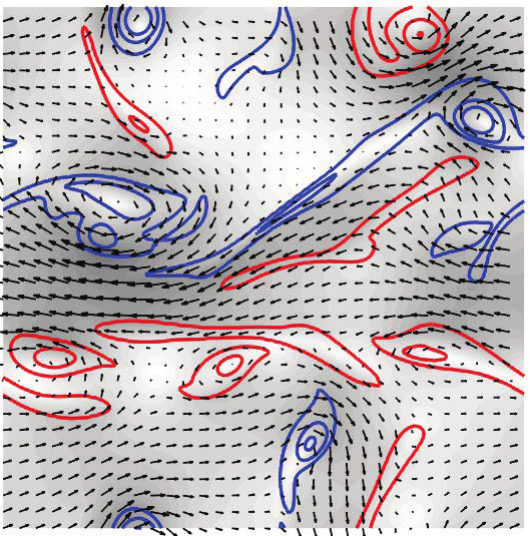}}% noploff256_t10_test7_tim10}}%
}%
\vspace*{3mm}%%%%%%%%%%%%%%%%%%%%%%%%%%%%%%%%%%%%%%%
\centerline{%
\raisebox{0mm}{\includegraphics[height=0.23\textwidth,clip]{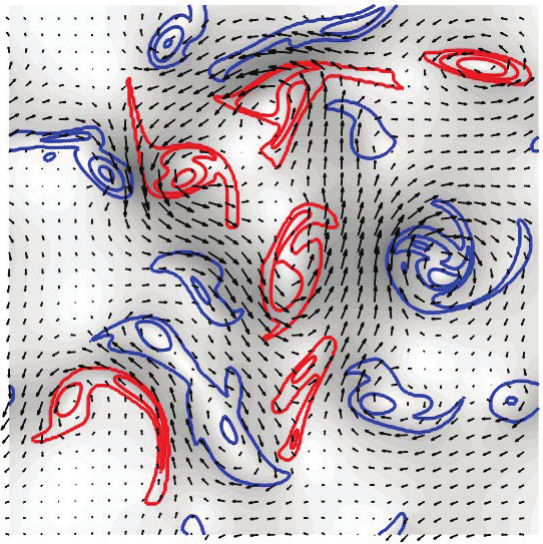}}% noploff256_t10_test4_tim0}}%
\mylab{-.27\textwidth}{.11\textwidth}{(c)}%
\hspace*{2mm}%
\raisebox{0mm}{\includegraphics[height=0.23\textwidth,clip]{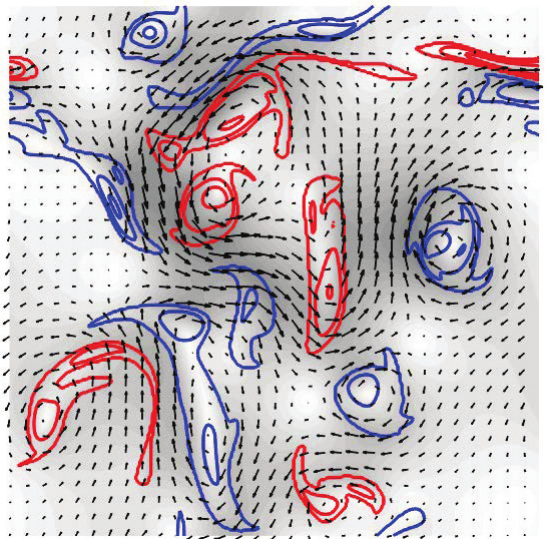}}% noploff256_t10_test4_tim3}}%
\hspace*{2mm}%
\raisebox{0mm}{\includegraphics[height=0.23\textwidth,clip]{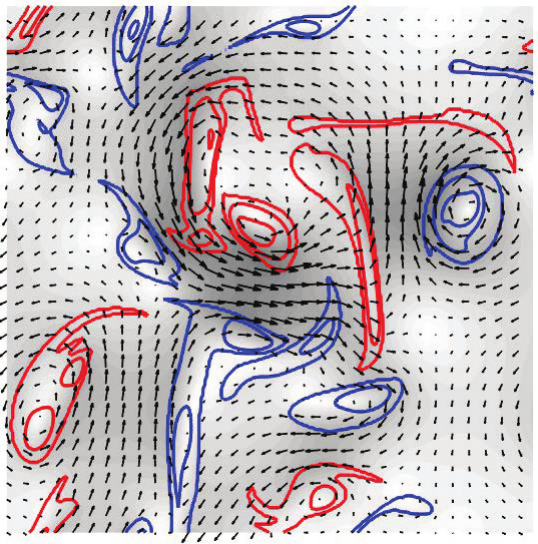}}% noploff256_t10_test4_tim7}}%
\hspace*{2mm}%
\raisebox{0mm}{\includegraphics[height=0.23\textwidth,clip]{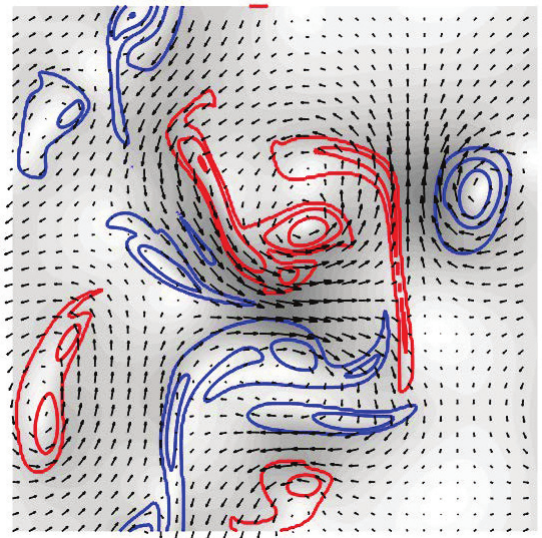}}% noploff256_t10_test4_tim10}}%
}%

\caption{%
Three examples of the organisation of vortices into streams. Time runs in each row from left
to right, $\omega_0't=0$, 2, 4.6, 6.5. Case T256. The three rows are independent
realisations, and only vortices with $s>\bra s \ket$ are included in the figure. Line
contours are positive (red) and negative (blue) vorticity, the arrows are velocity, and the
grey background is the velocity magnitude.
}
\label{fig:jets}
\end{figure}
% ===========================================================

% ------------------------------------------------------------------------------------------------------------
\subsection{The up-scale energy flux}\la{sec:tur2d}

The discussion in the previous section raises the question of how the dipoles get organised
into long streams. Three examples are given in figure \ref{fig:jets}. These examples
were chosen for didactic purposes to demonstrate the aggregation process, and are not truly
random. But they are fairly representative. Of the several thousand simulations available,
approximately 30\% were found visually to display an initial evolution similar to those in
the figure. Three of those were chosen at random from about 50 `good' cases. Each row in
the figure is a simulation, displayed at four approximately equidistant times, which are the
same for the three cases. Vortices are represented in colour, according to their sign, and
the arrows are the velocities. The intensity of the grey background is the velocity
magnitude. To best display the evolution, only the largest vortices $(s>\bra s\ket)$ are
included in each plot, and the Reynolds number is purposely chosen low.
 
Each simulation starts with a relatively disorganised arrangement of vortices but, at the
end of each sequence, positive (red) vortices have sorted themselves to one side of the
flow, and negative (blue) ones to the other, supporting a jet between them. Some
merging of like-signed vortices takes place in all cases. 

The question is how this happens, because, while continuity probably implies that the
velocity of any elongated velocity structure should be aligned with its axis, the opposite
is not true. Compact jet-like vortex dipoles with aspect ratios of order one (modons) are
well-known stable solutions of the Euler and Navier--Stokes equations
\citep{Flierl:80,mcwilliams80}.

Simulations of point vortex systems (not shown) spontaneously form tight dipoles and
corrotating pairs, but the dipoles do not organise into trains or jets, as is the case in figure
\ref{fig:jets}(a,b). Neither do they form stronger dipoles of vortex `clouds', as in figure
\ref{fig:jets}(c). When a point dipole collides with a solitary point vortex or with a corrotating
pair, it often loses one of its vortices, possibly breaking the colliding couple and forming a
new association. Very seldom the resulting arrangement involves more than two vortices.

Of course, sets of point vortices are Hamiltonian systems whose interactions conserve energy
\citep{bat67}. Both the merging of like-signed vortices and the breaking of an existing
dipole involve energy exchanges, which are much simpler if viscosity or filamentation can be
used as an energy dump. This would statistically favour the formation of the lower-energy dipoles, but any
such selection criterion requires a local mechanism to implement it. In particular, it is
unclear why a positive vortex being overtaken by a dipole would tend to reinforce the positive
component of the dipole, strengthening it, rather than merging with the negative one,
weakening it.

% ===========================================================
\begin{figure}
\vspace*{5mm}%
\centerline{%
\raisebox{0mm}{\includegraphics[height=0.40\textwidth,clip]{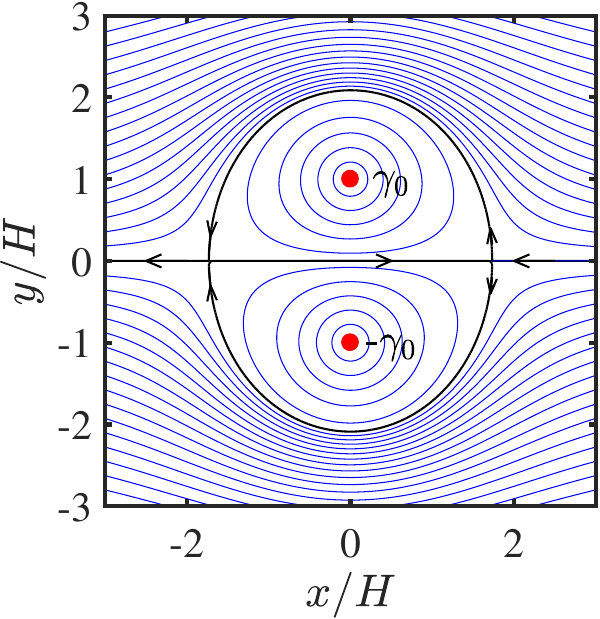}}%nakedipole}}%
\mylab{-.40\textwidth}{.35\textwidth}{(a)}%
\hspace*{5mm}%
\raisebox{0mm}{\includegraphics[height=0.40\textwidth,clip]{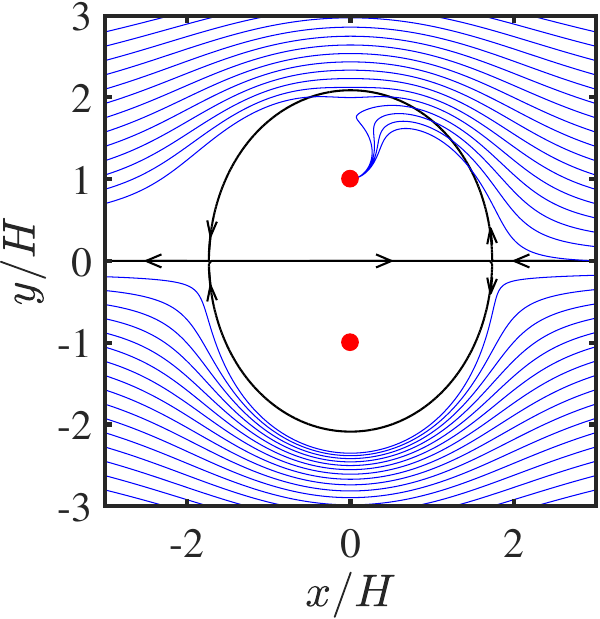}}%plusdipole}}%
\mylab{-.40\textwidth}{.35\textwidth}{(b)}%
}%
\caption{%
(a) Velocity field of a dipole of equal point vortices of circulation $\pm\gamma_0$ at distance
$2H$. The dipole is moving to the right, but is shown in the frame of reference linked to
the vortices. The blue lines are streamlines, as well as the trajectories of an advected
point vortex. The black lines are the dividing streamline in this frame of reference.
(b) As in (a), but the blue lines are the trajectories of the centre of gravity of a vortex patch of positive
circulation $\gamma$ and area $s$, such that $\gamma_0 s^2/\gamma H^4=2$. A patch with
negative circulation would be entrained to the negative vortex in the dipole. Only trajectories 
coming from large positive $x$ are included. See text and appendix \ref{sec:vpatch} for details.
}
\label{fig:drift}
\end{figure}
% ===========================================================

A possible mechanism is explained in figure \ref{fig:drift}. Consider the point-vortex
dipole in figure \ref{fig:drift}(a). In the comoving frame of reference, it forms a
recirculation bubble separated from infinity by an approximately elliptical dividing
streamline (see appendix \ref{sec:vpatch}). Any sufficiently weak point vortex
being overtaken by the dipole follows the streamlines around the bubble, independently of
its sign, and it is eventually left behind. There is no preference for which side of the dipole
its path takes, and it is therefore unlikely to statistically strengthen or weaken it.

The situation is different for the entrainment of an extended vortex, as detailed in
appendix \ref{sec:vpatch} for the case of a uniform vortex patch. Such patches drift with
respect to the advecting streamlines, as shown by the trajectories in figure
\ref{fig:drift}(b). Positive patches drift towards the positive component of the dipole, and
negative ones towards the negative component. The result is an average strengthening of the
dipole, and the formation of organised jets.
 
Note that this accretion model could explain the asymmetry observed in figure
\ref{fig:angleu}, where it was shown that the direction of the velocity of a stream is a
better predictor of the direction taken by the stream ahead than behind the position at which
it is measured. In the model presented here dipoles propagate forward, and take some time to
incorporate new vorticity. The jet of a growing dipole is strongest at its trailing edge,
and therefore predicts better the position of the leading dipole ahead of it.
Examples of this asymmetry between the front and back of dipoles can be seen in
figures \ref{fig:jets}(a) and \ref{fig:jets}(c), where a growing head dipole is followed by
a trailing jet of elongated filaments that have been created during their interaction with
the dipole.

On the other hand, although accretion explains why existing dipoles become stronger,
and thus how they segregate into classes, it does not explain how they organise into longer
streams. Several examples of both processes can be seen in figure \ref{fig:jets}. For
example, the flow in figure \ref{fig:jets}(c) organises into a single strong dipole but not
into a long stream. Figure \ref{fig:jets}(b) forms a horizontal leftwards jet in the
middle of the frame, but none of its dipoles is especially strong, and the same is true of
the oblique jet forming in the lower-right corner of figure \ref{fig:jets}(a). The growth of
the longitudinal scale in a set of two-dimensional vortices first came under
consideration when modelling the spreading of free shear layers in \cite{brownr}. Two
mechanisms were proposed at the time. In `vortex pairing', two vortices of the same sign merge into a larger
one \citep{win:bro:74}, while in `vortex tearing', a weaker vortex between two stronger ones
is strained into a filament that is eventually entrained or dissipated
\citep{moo:saff:75}. The accretion model developed above can be considered a case of
pairing, and a clear tearing can be seen between the second and third frames of figure
\ref{fig:jets}(b). Both processes are relevant in experimental shear layers
\citep{her:jim:82}, but that flow involves vorticity of a single sign, and the
scale growth mostly takes place along a single axis. Repeating a similar analysis in the
present mixed-sign isotropic situation would involve a more thorough processing than is possible
here, both theoretically and observationally.

The interaction of fast-moving dipoles with smaller vortices, akin to tearing, has been
invoked by \cite{dritsch08} as an important step of the forward enstrophy cascade. It takes
place when a third vortex is shredded by a dipole, but, to our knowledge, such interactions
have not been connected with the up-scale cascade. In a related observation, random forcing
of two-dimensional turbulence induces a self-similar enstrophy cascade, but the cascade
disappears (to form vortex crystals) when the forcing is partially deterministic, and the
character of the flow again changes when the forcing is fully deterministic, in which case
the flow becomes a `dilute gas of dipoles' that cleans most of the background low-level
vorticity \citep{jim:gueg:07}.

% ---------------------------------------------------------------------------------
\section{Discussion and conclusions}\la{sec:conc}

We have used simulation ensembles of decaying two-dimensional turbulence to study the early
stages of the evolution of the flow from a disorganised state towards a set of vortex cores
and large-scale structures of the kinetic energy. In this period, the dominant scale of the kinetic
energy is still small compared to the size of the computational box, and grows
monotonically. We have shown that, at least at the moderate Reynolds numbers of our
simulations, this growth is due to the appearance of elongated `streams' formed by a
concatenation of vortex dipoles. The growth of the energy scale is not due to the elongation
of the streams, whose aspect ratio stays in the range of 8--10, but to their proliferation,
and to the increase of the area fraction that they cover.

We have shown that the formation of the streams includes a process of aggregation of the
vortex cores. The cores segregate into two separate classes. Most of them are small and
mobile, and merge among themselves in a self-similar cascade that results in a power-law
probability distribution of vortex sizes \citep{Benzi92}. A few of the cores grow larger,
and eventually `freeze' into a low-mobility vortex system. These larger vortices are
responsible for most of the kinetic energy of the flow, but they themselves move slowly, in
what can be described as a quasi-equilibrium vortex `crystal'. The probability distribution
of their areas and circulations is exponential, rather than a power law, suggesting that
they do not grow by interacting among themselves, at least in the time scales considered
here, but by absorbing smaller vortices from the self-similar background. They are
responsible for the formation of the streams. We have proposed a formation mechanism by
noting that, although a vortex dipole shows no preference about how to entrain a point
vortex of either sign, the drift velocity of vortex patches biases positive patches to merge
with the positive component of the dipole, and negative ones to merge with the negative
component. As a consequence, the dipoles are strengthened and the streams are formed.

This growth mechanism of the energy scale is probably not the classical inverse
energy cascade, which is typically observed in forced, rather than in decaying flows. There
are two main properties that have to be explained for this cascade
\citep{tabeling02,Boff:Eck:12}. The first one is its $k^{-5/3}$ power spectrum, and the
second one is its low intermittency. The latter is consistent with the processes discussed
here, because growth by aggregation of small units of fixed size is an additive process,
which is not intermittent. But the observed exponential probability distribution of the
vortex size, and the approximately crystallisation, argue against the self-similarity implied
by a power spectrum. Note that there is no $k^{-5/3}$ plateau in the spectral slopes in
figure \ref{fig:specs}(c). It is possible that the aggregation of dipoles into streams is
hierarchical and self-similar in much larger simulations, but it is difficult to see how such a
self-similar amalgamation could avoid producing intermittency \citep{fsn78}. Moreover, the
facility with which streams are formed in figure \ref{fig:jets}, suggests that it would be
hard to prevent the large-scale vortex organisation from falling into local equilibrium. A
more appealing possibility is that the reason why a self-similar inverse cascade is only
observed in forced flows is that the effect of the forcing is to locally `melt' the vortex
crystal, much as a liquid develops short-range order and long-range disorder. Although
analysing such a model is beyond the scope of the present paper, a process of repeated
short-range crystallisation and longer-range melting caused by random excitations, is
probably not particularly intermittent. We mentioned above that partially deterministic
forcing can inhibit the forward enstrophy cascade \citep{jim:gueg:07}, but, to our
knowledge, no systematic study of the effect of forcing on the inverse cascade is available.

It is difficult not to be reminded by the discussion above of other examples of spontaneous
stream formation in more complicated flows. The best-known are probably the streaks in
wall-bounded turbulence and other shear flows \citep{tsuk:2006,dong17,jim18}, and the
azimuthal jets of planetary atmospheres and rotating flows
\citep{maltrud91,Drit:Mcint:08,Gross:ARFM:16,sacco19}. In many of these cases, the streams
are a streamwise concatenation of smaller units \citep{lozano-Q}, and the question arises of
how these units organise longitudinally. This is not the place to review the many models
proposed for this organisation, but most of them depend on the generation of new vorticity,
which is readily available from the shear or from the planetary rotation. There is no
vorticity generation in two-dimensional flow, and the mechanism discussed here, which
depends on the reorganisation of vorticity rather than on its creation, suggests that some
of these streams may, at least in part, share a common mechanism which is more related to
symmetry-breaking and pattern formation \citep{Cross:09} than to the dynamics of the
energy-generation process.
 
It is finally interesting to remark that, although the analysis in this paper is a fairly
classical example of hypotheses-driven research, it was made possible by following the
`blind suggestion', from the Monte--Carlo experiments in \cite{jotploff}, that dipoles are at
least as relevant to two-dimensional turbulence as individual vortex cores.
Hypothesis-driven science is, of course, the standard scientific method, and it can be
argued to be the only way in which the scientific corpus derives new `theories' from
the empirical accumulation of facts \citep{poincare08}. A different question is how
the hypotheses to be tested are chosen, and, preferably but not necessarily, whether the
resulting `theory' can be related back to this choice. In the present case, the original
intriguing observation was that dipoles of relatively small size had global effects
\citep{jotploff}, and it can easily be explained from the results of the subsequent analysis:
killing a dipole amounts to blocking a stream, and has effects over the length of the
stream. The mismatch between the size of the cause and of the effect is the elongation of the object
being modified. In a sense, the initial Monte-Carlo search is being used here as a
`hypothesis generator', while the rest of the paper is the `hypothesis test', and it is tempting
to speculate that this could be a reasonable division of labour between computer and
researcher. Further discussion of these `epistemological' issues can be found in
\cite{jimploff20,jotploff}. 

%%%%%%%%%%%%%%%%%%%%%%%%%%%%%%%%%%%%%%%
\vspace{2ex}
This work was supported by the European Research Council under the Coturb grant
ERC-2014.AdG-669505. 
%

% ----------------------------------------------------------------------------------------------------
%\bibliographystyle{jfm}
%\bibliography{cit2}

%----------------------------------------------------------------------------------------------------------------------------
\appendix
\section{Poisson distribution of the closest point}\la{sec:poisson}

If the expected number of Poisson-distributed points in a set parametrised with $r$ is
$\lambda(r)$, the probability of finding no points within the set is
\beq
P_0(r) = \exp(-\lambda), 
\la{eq:poiss0}
\eeq
and the probability density that the first point is precisely at $r$ is
\beq
P_c(r) = -(\dr P_0/\dr r) = (\dr \lambda/\dr r) \exp(-\lambda). 
\la{eq:poiss1}
\eeq
For a set of points distributed in a plane with uniform average density $\rho$, the expected
number of points within a distance $r$ of a reference position is $\lambda=\pi \rho r^2$,
and
\beq
P_c(r) = 2\pi\rho r \exp(-\rho \pi r^2 ).
\la{eq:poiss2}
\eeq
This distribution can be integrated to show that the average distance to the closest point is $\bra r\ket_0 =
1/\sqrt{4\rho}$, so that \r{eq:poiss2} can be written as
\beq
P_c(\xi) = (\pi\xi/2)  \exp(-\pi \xi^2/4),
\la{eq:poiss3}
\eeq
where $\xi=r/\bra r \ket_0$. 

If we add the restriction that no point can be closer than a distance $a$ (e.g. the vortex diameter), the expected number of points within a distance $r\ge a$ is $\lambda=\pi\rho (r^2-a^2)$, 
\beq
P_c(r) = 2\pi\rho r \exp[-\rho \pi (r^2-a^2) ],
\la{eq:poiss1a}
\eeq
and the average distance to the closest point becomes
\beq
\bra r\ket_\epsilon = \int_a^\infty r P_c(r) \dd r = a +(1/2\sqrt{\rho})\, \mathrm{erfc} (\epsilon) \exp(\epsilon^2),
\la{eq:poiss2a}
\eeq
where $\epsilon=a/\sqrt{\rho\pi}=\sqrt{4/\pi}\,a/\bra r \ket_0$. It can be shown that $\bra
r\ket_\epsilon\approx a$ for $\epsilon\gg 1$, but that $\bra r\ket_\epsilon\approx \bra
r\ket_0$ for $\epsilon\lesssim 0.5$. Figure \ref{fig:vortices}(e) suggests that
$\widehat{a}=a/\bra r \ket_0\approx 0.2$--0.3, so that
\beq
P_c(\xi) \approx (\pi\xi/2)  \exp[-\pi (\xi^2-\widehat{a}^2)/4],\quad \xi>\widehat{a}.
\la{eq:poiss3a}
\eeq
The effect is to crop the part of the distribution near the origin, while raising its peak to
compensate for the missing mass (as in figure \ref{fig:specs}e).

% -----------------------------------------------------------------------------------------------
\section{Drift of a vortex patch.}\la{sec:vpatch}

While point vortices are advected by the flow velocity, vortices with a wider support drift
with respect to it. In the particular case of small patches of uniform vorticity, the drift
velocity can be computed as a series expansion of the vortex radius. The following equations of motion
are drawn from \cite{jimenez88}.

Define a complex variable $z=(x+\ii y)/H$, where $H$ is a characteristic length scale, and
consider a uniform vortex patch of circulation $\gamma\Gamma$ and area $s$. Using $\Gamma$ and $H$
to define the time and length scales, the expansion parameter is $\epsilon^2=s/\pi H^2$,
which is assumed to be small. The irrotational complex flow velocity in the absence of the
patch is described by an analytic function $w_\infty(z)= (u-\ii v)\,H/\Gamma$, with a
similar non-analytic expression within the patch. To lowest order, the contour of the patch
is an ellipse,
\beq
z-z_c = \epsilon\eta (1+b_2 \epsilon^2/\eta^2),
\la{eq:patch1}
\eeq
where $z_c$ is the centre of gravity of the patch, and $\eta=\exp(\ii \phi)$ is the unit
circle. Matching at this contour the expansions of the velocity inside and outside the patch
provides an evolution equation for the ellipticity,
\beq
2\pi\epsilon^2 \dd b_2/\dr \tau = \ii \gamma b_2 +c_2^*,
\la{eq:patch2}
\eeq
where $\tau=\Gamma t/H^2$ is a rescaled time, the asterisk stand for complex conjugation,
and
\beq
c_k = \frac{2\pi}{(k-1)!}\,\frac{\dr^{k-1} w_\infty}{\dr z^{k-1}}(z_c).
\la{eq:patch3}
\eeq
The drift velocity, defined as $\dr z^*_c/\dr \tau = w_\infty(z_c)+w_d$, can be expressed as
\beq
w_d= (\epsilon^4/2\pi) b_2 c_3,
\la{eq:patch4}
\eeq
and, if we further assume that \r{eq:patch2} has reached equilibrium, so that $b_2= \ii c_2^*/\gamma$, 
\beq
w_d = \frac{\pi\ii \epsilon^4 }{\gamma}\,\frac{\dr w^*_\infty}{\dr z^*}\,  \frac{\dr^2 w_\infty}{\dr z^2}. 
\la{eq:patch5}
\eeq
Consider now the effect on the patch from a dipole formed by two point vortices of circulation
$\Gamma\gamma_0$ separated by a distance $2H$. In figure \ref{fig:drift}(a) in the body of the
paper, the positive vortex is on top, and the dipole would move to the right, but it is made
stationary by a uniform negative velocity at infinity. Consider a third vortex being
overtaken by the dipole, but neglect its effect on the dipole itself. In the units defined
above, the velocity induced by the dipole is
\beq
w_\infty = \frac{\gamma_0}{\pi} \left( \frac{1}{1+z^2}-\frac{1}{4}\right). 
\la{eq:patch6}
\eeq
Its effect on a point vortex is given by the streamlines in figure \ref{fig:drift}(a). The
vortex is deflected around the recirculation bubble of the dipole, and eventually left
behind. There is no difference between a positive and a negative vortex, but the situation
is different for an extended patch, because the drift velocity,
\beq
w_d = - \frac{4\ii\gamma_0^2\epsilon^4}{\gamma\pi}\, \frac{z^*}{1+z^{*2}}\,\frac{3z^2-1}{(1+z^2)^3}, 
\la{eq:patch7}
\eeq
depends on the sign of the circulation of the patch being overtaken. As shown in figure \ref{fig:drift}(b),
positive vortices tend to be entrained into the upper part of the stream, and to merge with
the positive vortex of the dipole. Negative patches are entrained towards the lower negative
vortex. The result is that the dipole is reinforced on average.

\end{document}